\documentclass[conference]{IEEEtran}
\IEEEoverridecommandlockouts
\usepackage[hidelinks]{hyperref} 
\hypersetup{ 
    colorlinks=false, 
    pdftitle={Consumer Group Autoscaler}
}
\usepackage{cite}
\usepackage{caption}
\usepackage{amsmath,amssymb,amsfonts}
\usepackage{float}
\usepackage[caption=false, font=footnotesize]{subfig}
\usepackage{algorithm, algorithmic}
\usepackage{mathtools}
\usepackage{xpatch}
\usepackage{xcolor}
\usepackage{listings}
\usepackage{realboxes}
\usepackage{verbatim}
\usepackage{bm}
\definecolor{codegreen}{rgb}{0,0.6,0}
\definecolor{codegray}{rgb}{0.5,0.5,0.5}
\definecolor{codepurple}{rgb}{0.58,0,0.82}
\definecolor{backcolour}{rgb}{0.95,0.95,0.92}
\lstdefinestyle{mystyle}{
    backgroundcolor=\color{backcolour},   
    commentstyle=\color{codegreen},
    keywordstyle=\color{magenta},
    numberstyle=\tiny\color{codegray},
    stringstyle=\color{codepurple},
    basicstyle=\ttfamily\footnotesize,
    breakatwhitespace=false,         
    breaklines=true,                 
    captionpos=b,                    
    keepspaces=true,
    numbers=left,                    
    numbersep=5pt,                  
    showspaces=false,                
    showstringspaces=false,
    showtabs=false,                  
    tabsize=2
}
\makeatletter
\xpretocmd\lstinline{\bgroup\appto\lst@DeInit{\egroup}}{}{}
\makeatother
\usepackage[pdftex]{graphicx}
\usepackage{textcomp}
\usepackage{xcolor}
\def\BibTeX{{\rm B\kern-.05em{\sc i\kern-.025em b}\kern-.08em
    T\kern-.1667em\lower.7ex\hbox{E}\kern-.125emX}}
\begin{document}

\title{
    Multi-Objective Optimization of Consumer Group Autoscaling in Message Broker Systems
    }

\author{ 
    \IEEEauthorblockN{
        Diogo Landau\IEEEauthorrefmark{1},
        Nishant Saurabh\IEEEauthorrefmark{1},
		Xavier Andrade\IEEEauthorrefmark{3}        
        Jorge G Barbosa\IEEEauthorrefmark{2}\\
    }
    \IEEEauthorblockA{
        \small
        \IEEEauthorrefmark{1}Department of Information and Computing Sciences, Utrecht University, NL\\
        \IEEEauthorrefmark{3}Department of Industrial Engineering, University of Porto - FEUP, Portugal\\       
        \IEEEauthorrefmark{2}LIACC, Faculdade de Engenharia,  Universidade do Porto, Portugal\\
        \IEEEauthorrefmark{1}d.landau@uu.nl,
        \IEEEauthorrefmark{1}n.saurabh@uu.nl,
        \IEEEauthorrefmark{3}xavier.andrade@fe.up.pt,
        \IEEEauthorrefmark{2}j.barbosa@fe.up.pt
    }
}

    
    
    

\maketitle
\begin{abstract}
    Message brokers often mediate communication between data producers and
    consumers by adding variable-sized messages to ordered distributed queues.
    Our goal is to determine the number of consumers and consumer-partition
    assignments needed to ensure that the rate of data consumption keeps up
    with the rate of data production. We model the problem as a variable item
    size bin packing problem. As the rate of production varies, new
    consumer-partition assignments are computed, which may require rebalancing
    a partition from one consumer to another. While rebalancing a queue, the
    data being produced into the queue is not read leading to additional
    latency costs. As such, we focus on the multi-objective optimization cost
    of minimizing both the number of consumers and queue migrations. We present
    a variety of algorithms and compare them to established bin packing
    heuristics for this application. 
    Comparing our proposed consumer group assignment strategy with Kafka's, a
    commonly employed strategy, our strategy presents a $90^{th}$ percentile
    latency of $4.52s$ compared to Kafka's $217s$ with both using the same
    amount of consumers. Kafka's assignment strategy only improved the consumer
    group's performance with regards to latency with configurations that used
    at least $60\%$ more resources than our approach.
\end{abstract}

\maketitle
\section{Introduction}

\IEEEPARstart{B}{rokers} are a common tool employed to fixture communication in
a distributed environment \cite{magnoni2015modern}. This system  provides with
asynchronous communication between producers and consumers, and handles some of
the challenges that are common within distributed and concurrent data
processing \cite{john2017survey}. 

We aim to solve the consumer group autoscaling problem for a generalized
message broker use, where each queue has an independent production rate, and
the messages published present variable sizes. When solving for the
consumer-partition assignment, the first concern is to minimize the number of
consumers required, which relates to an operational cost. A second cost, is a
direct consequence of a message broker's load varying over time. This implies
that to comply with service level agreements, a partition might need to migrate
from one consumer to another. While migrating a partition its data is not
consumed, hence a rebalance/migration cost related to latency SLA violations.
The conflicting objectives of minimizing the operational and rebalance costs,
make this consumer group autoscaling problem multi-objective.

Any service that reads data from a message broker via a group of consumers
would benefit from a solution that elastically scales the consumers based on
the current demand. This would allow reducing both the operatinal cost and the
latency SLA violations.

However, existing approaches attempting to provide such a solution fall short
as they do not consider skewed distributions in terms of the load of each queue
\cite{chindanonda2020self}, or they are only adequately modelled in scenarios
where the message size is constant \cite{ezzeddine2022cost}.

We consider a queue or partition to be a structure within a message broker
environment where messages are appended in the same order as they were produced
\cite{raynal2013distributed, lumezanu2006decentralized}. It is also important
that a queue is capable of delivering messages in the same order as they were
produced. This consumption model is often a requirement to guarantee state
consistency between two distinct distributed services for a specific business
related entity. Common applications that read the same set of messages in their
production order include event-carried state transfer (a pattern which is
commonly used in microservice arcthitectures), and system state simulation. The
former consists of having two distributed services reading the same set of
messages so as to replicate an entitie's state within their data store
\cite{burckhardt2012eventually, su2005slingshot}, whereas the latter is used to
simulate a system's state at a given point in time by reproducing the messages
up until that point \cite{carbone2017state, kshemkalyani1995introduction}.

Two common message broker implementations, Kafka\footnote{Kafka Message
Ordering, \url{https://kafka.apache.org/}} and RabbitMQ\footnote{RabbitMQ
Message Ordering:
\url{https://www.rabbitmq.com/queues.html\#message-ordering}}, have different
ways of guaranteeing message order on consumption. Within Kafka, all the
messages produced to a partition are appended and delivered in the same order
as they were published. When consuming the data, only a single consumer in a
group can be assigned a partition, and therefore message ordering is guaranteed
for a single partition. As for RabbitMQ, it guarantees messages toward a queue
are enqueued in the same order as they were produced. Therefore, to guarantee
same order message delivery, only a single consumer can be reading messages
from the queue.

Considering consumers as bins and queues as items that have to be assigned a
bin, this problem was modeled as a variation of the Bin Packing (BP) Problem.
Since the rate at which data is inserted into a queue varies over time, a BP
item's size also varies with time. In fact, a queue's size correlates to its
current write speed, which fluctuates based on the current system's load. This
inevitably implies that a BP solution for a given time instant may not hold
true in future instants.

On account of this BP variation, a new solution has to be computed at different
time instants, each having different information as to the size of each item
(each queue's write speed for that time instant). This might lead to a queue
(item) being migrated to a different consumer (bin) when compared to the
consumer group's previous configuration. 

Since two consumers cannot read from the same queue concurrently, there is
another cost to take into account associated to rebalancing a partition
\cite{van2021performance}. This cost is related to the amount of data that is
not being read while the queue is being rebalanced.

Existing heuristic bin packing algorithms do not take this rebalance cost into
account, which leads to a higher number of migrations and consequently to
higher migration costs. Hence, we propose four new BP heuristic algorithms that
account for the rebalance costs, three of which are shown in Section
\ref{sub:exp_modified_any_fit} to be a competitive alternative to the
multi-objective optimization problem at hand. We also propose an Rscore in Sec.
\ref{sub:rscore} metric to evaluate an algorithm's rebalance cost for each
iteration. 

This problem becomes more relevant within skewed distributed event queues
\cite{ezzeddine2022cost}, wherein the rate of production into each queue is
independent, and not necessarily equal for the different queues, i.e., the
event queues will present different concentrations of messages. In fact, common
message broker implementations use an event's key to deem the queue it is to be
inserted in. Keys used with more frequency than others are the cause of this
uneven distribution of events in the different queues.

The contributions we present in this paper are: 
\begin{itemize}

    \item Modeling the consumer group autoscaling problem for a generalized
    message broker use; 
    
    \item The Rscore metric, which quantifies a rebalance cost between two
    consecutive consumer group assignments; 
    
    \item Four modified approximation algorithms;
    
    \item An application of the proposed model and algorithms in a
    \textit{Kafka} message broker environment; 
    
    \item Evaluating the autoscaler implementation's response time when
    autoscaling the consumer group in a production infrastructure.

\end{itemize}

In Section \ref{sec:related_work}, we present applications of the Bin Packing
Problem, and a comparison of existing approximation algorithms to solve the Bin
Packing problem. Section \ref{sec:problem_formulation}, formally defines the
problem, followed by the proposed approach in Section \ref{sec:algorithm}. We
also present in Section \ref{sec:autoscaler} an application of autoscaling a
group of consumers based on the broker's current load within Kafka. Section
\ref{sec:experimentation} evaluates the proposed algorithms, and the
autoscaling model when applied to a production Kafka environment. Lastly, we
conclude this paper in Section \ref{sec:conclusions}.

\section{Related Work}
\label{sec:related_work}

This section starts by presenting in which other contexts the bin packing model
has been applied, to demonstrate its effectiveness when representing a
minimization and assignment problem similar to the one being studied.

Moreover, due to bin packing's NP-hard nature, and the real-time constraints of
autoscaling a consumer group, we present existing heuristic algorithms that
solve this optimization problem in polynomial time. These algorithms are also
used as a comparison to our proposed algorithms in Section
\ref{sec:experimentation}.

Lastly, we present the current state of consumer group autoscaling. Despite the
the fact that the algorithms presented attempt to solve the same problem, some
of their assumptions (e.g. equal production rate into queus; all messages
produced having constant size) limit their applicability in most production
message broker environments. Given the fact that our algorithms are not based
on the same assumptions, these cannot be used for comparison without some
adjustments.

\subsection{Bin Packing Applications}

There are several applications where BP is used to provide an optimal or
sub-optimal solution \cite{nardelli2019efficient, FURINI2012251,
dell2020branch, delorme2017logic, xia2013throughput, gulisano2012streamcloud,
heinze2013elastic}. The Virtual Machine Placement (VMP) problem, has gained
more attention due to the increasing use of cloud providers to support
companies' technological infrastructure. The problem can be generalized to a
set of Virtual Machines (VM), each having to be assigned one of the cloud
provider's Physical Machines (PM) while attempting to minimize the operational
cost (number of PMs used). 

As new VM requests arrive, each new item (VM) has to be assigned a bin (PM).
Furthermore, an item that has been previously assigned can be migrated to
another physical machine since migration is made available via virtualization
technology. Therefore, VMP is considered a fully dynamic bin packing problem
where items arrive and depart over time (online bin packing) and migrations are
allowed \cite{kamali2016efficient, shi2013empirical, mann2015approximability,
song2013adaptive}. Moreover, the migration cost at each decision interval is
measured either as the number of migrations \cite{shi2013empirical} or is
computed based on the size of all VMs that have to be migrated
\cite{mann2015approximability}.

Song et al. \cite{song2013adaptive} and Kamali et al.
\cite{kamali2016efficient} further investigate an additional case wherein the
VMs (items) require a variable amount of resources throughout their lifetime,
and therefore the size of the items varies over time. This constitutes the
variable item size bin packing problem (VISBP), which is the same problem we
are studying in the consumer group autoscaling context.

\subsection{Approximation Algorithms} 
\label{sub:approximation_algorithms}

The Bin Packing (BP) problem is a well established research problem, and has
been extensively reviewed in the literature \cite{hadary2020protean,
panigrahy2011heuristics, berndt2020fully}. Due to the time constraints imposed
by our application, this paper gives emphasis to the Approximation Algorithms
that heuristically solve the problem in low-order polynomial time, as opposed
to the higher time complexity Linear Programming approach
\cite{fleszar2002new}.

A method is presented in \cite{coffman2013bin} to classify the BP problem,
which will be used throughout this section. During an algorithm's execution, a
bin can find itself either \textit{open} or \textit{closed}. In the former, the
bin can still be used to add additional items, whereas in the latter, it is no
longer available and has already been used.  

The list of bins is indexed from left to right, and the number of bins used by
an algorithm can be computed using the index of the first empty bin in the list
of bins. When creating a bin, this process can be visualized as opening the
lowest index of the empty bins (left-most empty bin).

Using $A(L)$ to denote the amount of bins a certain algorithm makes use of for
a configuration of items $L$, $OPT(L)$ to represent the amount of bins required
to achieve the optimal solution, and defining $\Omega$ as the set of all
possible lists, each with a different arrangement of their items, $R_A(k)$ (Eq.
\ref{eq:performance_ration}) encodes a performance metric relative to the
optimal number of bins used by an algorithm.

\begin{equation}
    \label{eq:performance_ration}
    R_A (k) = \sup_{L \in \Omega} \left \{ \frac{A(L)}{k} : k = OPT(L) \right \}.
\end{equation}

The Asymptotic Performance Ratio (APR) of an algorithm $A$ ($R_A^\infty$) is
defined by Eq. \ref{eq:APR}, and will be used throughout this section to
compare the algorithms' performance \cite{coffman2013bin}.

\begin{equation}
    \label{eq:APR}
    R_A^\infty = \limsup_{k \to \infty} R_A(k).
\end{equation}

The input to the consumer group autoscaling problem is a list of the production
rates into each queue. Therefore, the queue-consumer mapping can be performed
offline as all items' sizes are known beforehand. However, online algorithms
can still be used, albeit with an increase in operational cost compared to
their offline counterparts. As a result, out of the set of classes defined in
\cite{coffman2013bin}, the ones which are of interest for this paper, are the
Offline and Any Fit algorithms. We extend the Any Fit algorithms in Sec.
\ref{sec:algorithm} as part of our methodology, and we adapt the Offline
algorithms to be applicable in this multi-objective context to serve as a
baseline with which to compare the performance of our proposed algorithms in
Sec. \ref{sec:experimentation}.

An algorithm belonging to the Any Fit class of algorithms must satisfy the
following conditions \cite{coffman2013bin}: \textit{if bin $j$ is empty, an
item will not be assigned to it if the item fits into any bin to the left of
$j$}. As shown by Coffman et al. \cite{coffman1999bin}, any online heuristic
that fits these constraints shall have an APR as between $\frac{17}{10}$ and
$2$. 

The Any Fit class of algorithms perform best if the list of items is sorted in
decreasing order prior to assigning items to the bins. The following offline
approximation algorithms, apply this sorting strategy, and all but the Next Fit
Decreasing (NFD) belong to the aforementioned class. The First Fit Decreasing
(FFD) places each item in the left-most bin as long as the bin's capacity is
not exceeded. As shown in \cite{johnson1974worst}, this algorithm's APR is
$R_{FFD}^\infty = \frac{11}{9}$

For each item, The Worst Fit Decreasing (WFD) attempts to assign an item to the
the existing bin with most slack. If the size constraint is not satisfied, a
new bin is created and assigned. This algorithm has as APR $R_{WFD}^\infty
\approx 1.6910$. Similarly, the Best Fit Decreasing (BFD) differs only on its
packing strategy, wherein it attempts to place the item in the bin where it
fits the tightest. In case there is no open bin to fit the current item, a new
bin is created where the item is inserted \cite{man1996approximation}. The BFD
heuristic has an APR of $R_{BFD}^\infty = \frac{11}{9}$.

Lastly, the NFD only has a single open bin at a time, which is also the last
created bin. This algorithm attempts to place the item in the right-most
non-empty bin, and, if it doesn't fit, then the bin to the right is used. As shown
in \cite{baker1981tight}, this algorithm has as APR $R_{NFD}^\infty \approx
1.6910$.

\subsection{Consumer Group Autoscaling}

To comply with latency service level agreements in a publish subscribe message
broker environment, it is common to provide with consumer group parallelism to
split the load between the elements of a group \cite{kamburugamuve2016survey,
thein2014apache, kamburugamuve2015framework}. As such, when the rate of
production increases and the consumer group is falling behind, it is expected
that the group of consumers upscales and reassigns the partitions.

Currently, Kubernetes Horizontal Pod
Autoscaler\footnote{https://kubernetes.io/docs/tasks/run-application/horizontal-pod-autoscale/}
(HPA) paired with KEDA\footnote{https://keda.sh/docs/2.9/scalers/apache-kafka/}
enables consumer group autoscaling based on average CPU, memory and event lag
between the consumers of the group. These threshold-based techniques scale the
consumer group incrementally, depending on the group's current average
performance with regard to the chosen strategy. Chindanonda et al.
\cite{chindanonda2020self} devise a strategy to forecast the rate of
production, which then enables computing an estimated number of consumers
required in the consumer group. Given a set of consumers, the threshold-based
strategies (e.g. HPA; KEDA) and the algorithm presented in
\cite{chindanonda2020self} distribute the number of partitions evenly between
the consumers. This is also the default strategy used by the state-of-the-art
message broker \textit{Kafka}. This strategy fails to load balance between the
different consumers when the workload presents a skewed distribution, since the
production into each event queue is independent and not necessarily equal.

Ezzeddine et al. \cite{ezzeddine2022cost} emphasize the skewed distribution
problem, and resort to the consumer group lag relative to each queue to monitor
the event distribution. Given a consumer's capacity and the lag of each queue,
the authors compute the partition to consumer assignments through the First Fit
Decreasing Bin Packing algorithm. Although the authors present a load-aware
partition-consumer assignment, using the lag as a metric to determine each
queue's production rate is inaccurate when the messages published to each queue
have variable sizes. Since the consumption rate is limited by the network
throughput, to improve a queue's load model, the production and consumption
rates should be measured in bytes per second ($bytes/s$). Since we study the
case where messages have variable sizes, the approach provided by
\cite{ezzeddine2022cost} is not directly applicable to our problem, so we modify
their approach by measuring the production and consumption rates in $bytes/s$
and use it to compare with our proposed algorithms.

The cost of rebalancing partitions between the consumers due to data
consumption downtime from the partitions being rebalanced, is also highlighted
in \cite{chindanonda2020self}. Despite this fact, the only mitigation strategy
presented was to impose a time constraint as to when the group could scale
based on its last scaling decision. With regards to \cite{ezzeddine2022cost},
the authors do not take the rebalance cost into account.

\subsection{Summary}

The consumer group autoscaling state of the art focuses on solving for
distributed queue scenarios which rarely apply in real world production
environments. The first common assumption, is that the production rate into all
distributed queues is equal (attempting to assign the same number of partitions
to the different consumers in a group), whereas the second, is that all
messages produced into the different event queues have the same size (measuring
production and consumption rates in number of events per second). Existing
solutions also lack a deterministic metric to account for the rebalancing cost,
and present conservative heuristic algorithms to migrate the items to different
bins (in some cases items with higher cost are not migrated). 

For these reasons, we propose four heuristic algorithms that simultaneously aim
to minimize the operational and migration costs in Sec.
\ref{sub:modified_any_fit}, and we provide an Rscore metric in Sec.
\ref{sub:rscore} to evaluate the rebalance/migration cost.

\section{Problem Formulation}

\label{sec:problem_formulation}

\begin{figure}[htb!] 
    \centering
    \includegraphics[width=0.4\textwidth]{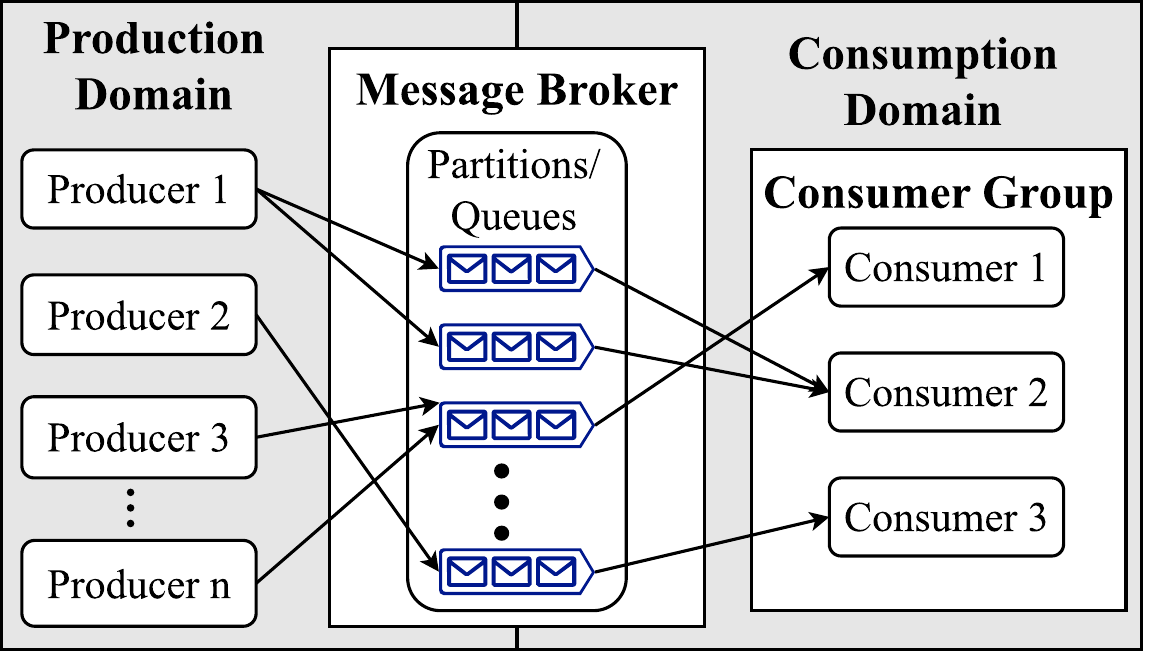} 
    \caption{
        Representation of data production and consumption domains within a
        message broker environment. 
    } 
    \label{fig:kafka_representation} 
\end{figure}

A group of consumers is interested in consuming data from several
partitions/queues with varying load within the message broker. The set of
partitions is assigned to the different consumers of the group to
simultaneously parallelize and load-balance data consumption. This problem is
depicted in the Data Consumption Domain of Figure
\ref{fig:kafka_representation}. 

The aim of our work is to achieve a deterministic approach to determine the
number of consumers required working in parallel, so as to guarantee that the
rate of production into a partition is not higher than its rate of consumption
by the consumer group, while minimizing the operational cost. In contrast, if
this were not the case, messages would accumulate leading to the lag between
the last message inserted and the last message read by the consumer group to
increase with time. Additionally, to make sure the messages are consumed in the
same sequence as they were published, a single partition can only be assigned a
single consumer. We also assume that each consumer in the group presents an
equal maximum consumption rate.

The model for this problem fits the constraints of the single bin size Bin
Packing problem, where the bins are consumers that have as capacity their
maximum consumption rate, and the weights are the partitions and their
respective write speeds. The problem then is to find the minimum amount of
consumers where to fit all the partitions so as to make sure that the sum of
the write speeds of the partitions assigned to a single consumer does not
exceed its maximum capacity $C$. 

Given a measurement of the partitions' write rate at time instant $t$, the
problem is of minimizing the operational cost (number of consumers), while
guaranteeing that each partition is assigned to a single consumer. As such, we
use the Bin Packing formulation (Eq. \ref{BPP model} - \ref{opt:c4}) with the
time variable $t$ which accounts for the time at which the measurement and the
new consumer group configuration is being computed.

\begin{alignat}{3}
\label{BPP model}
    &\min
        &&\sum_{i \in B} y_i(t) 
            && \\
    &\text{subject to} \quad
        && \sum_{j \in P} \frac{s_j(t)}{C} \cdot x_{ij}(t) \leq y_i(t) \quad
            && \forall \; i \in B \label{opt:c1}\\
    &   && \sum_{i \in B} x_{ij}(t) = 1, \quad
            && \forall \; j \in P \label{opt:c2}\\
    &   && y_i(t) \in \{0, 1\}
            && \forall \; i \in B \\
    &   && x_{ij}(t) \in \{0,1\}
            && \forall \; i \in B, j \in P. \label{opt:c4}
\end{alignat}
Set $B$ corresponds to the available consumers, and $P$ as the list of
partitions to be arranged into consumers. Decision variable $x_{ij}(t)$ and
$y_i(t)$, indicate whether a partition $j$ is assigned to a consumer $i$, and
whether consumer $i$ is used, respectively, at a time instant $t$.
Additionally, $x_{ij}(t)$ constructs a matrix $\bm{\mathrm{X}}(t)^{|B| \times
|P|}$ and $y_i(t)$ provides the column vector $\bm{\mathrm{y}}(t)^{|B| \times
1}$. The production rate measurement of a partition $p$ at time $t$ in
$bytes/s$ is denoted by $s_j(t)$, and $C$ denotes the maximum consumption rate
of a consumer. As for Constraints \ref{opt:c1} and \ref{opt:c2}, the former
assures that the sum of the partitions' write speed assigned to any consumer
does not exceed its capacity, and the latter guarantees that each partition is
assigned a consumer.

Hitherto, we have only considered the bin packing formulation for a single
iteration, nevertheless we must adapt the formulation to the dynamicity present
due to the production rate into each queue varying over time $t$. As a result, we
arrive at the Variable Item Size Bin Packing (VISBP) formulation, also
considered in \cite{song2013adaptive} and \cite{kamali2016efficient}.

A VISBP heuristic algorithm is executed sequentially and at discrete time
instants. We use $k, k \in \mathbb{N}$, to denote the sequential iterations at
which a new consumer group configuration is computed. Additionally, $t_k$
represents an iteration's execution time instant. It is also important to note
that between any two consecutive iterations $k$ and $k-1$, $t_k - t_{k-1}$ is
variable. This occurs because a new partition to consumer assignment is
computed depending on a set of conditions, e.g., whether a single consumer has
its capacity exceeded or whether a certain amount of time has exceeded since
the last time a reconfiguration was evaluated. This is further described in the
state machine presented in Sec. \ref{autoscaler:controller}.

At any time $t < t_1$, the consumer group is not yet assigned any partitions
and therefore: 
\begin{equation}
    \bm{\mathrm{X}}(t) = 
        \begin{bmatrix}
            0 & ... & 0 \\
            \vdots & \ddots & \vdots \\
            0 & ... & 0
        \end{bmatrix}, 
        \forall \; t < t_1
\end{equation} 

Also, for any given $t_k, k > 1$, there already exists a partition-consumer
assignment computed at $t_{k-1}$, $\bm{\mathrm{X}}(t_{k-1})$. Then, we compute
$\bm{\mathrm{\Delta}}(k)$ to determine which partitions have to be reassigned
to different consumers in Eq. \ref{eq:delta}.

\begin{equation}
\label{eq:delta}
    \bm{\mathrm{\Delta}}(k) = \bm{\mathrm{X}}(t_k) - \bm{\mathrm{X}}(t_{k-1}) \quad \forall \; k > 1,
\end{equation}

For a cell $\delta_{ij}(k)$ in $\bm{\mathrm{\Delta}}(k)$, if its value is -1 it
indicates that the consumer $i$ should stop consuming from partition $j$. If
$\delta_{ij}(k) = 1$, then consumer $i$ should start consuming from partition
$j$. Lastly, if the value is $0$, then no change is made in consumer $i$'s
assignment with regards to partition $j$. 

Given that $\bm{\mathrm{\Delta}}(k)$ encodes which partitions have to be
reassigned from one consumer to another, there is an additional cost to take
into account related to rebalancing a partition. Only one consumer can be
reading from a partition at a time, and therefore, when a partition is to be
reassigned, that is assigning a partition to another consumer, the one it is
currently assigned to has to stop consuming in order to allow the new consumer
to start. Due to this process, there is some downtime where data is not being
consumed from the partition being migrated. 

As such, the rebalance cost has to increase with the write rate of the
partitions being migrated, i.e. as the write rate of a partition being
reassigned increases, the rate at which unread data accumulates during the
rebalance phase, also increases. 

\section{Proposed Algorithm}
\label{sec:algorithm}

\subsection{Rscore}
\label{sub:rscore}

Within this problem's context, rebalancing inevitably implies having the
consumer group stop consuming from a partition while it is being reassigned
from one consumer to another. There cannot be a concurrent read from the same
partition by members of the same group.

This paper introduces the Rscore metric to compute the total rebalance cost of
a group's reconfiguration. The metric reflects the impact of stopping the data
consumption from the partitions that are being rebalanced. 

A migration is evaluated between two consecutive iterations and, therefore, a
partition requires rebalancing at iteration $k$ if its column in
$\bm{\mathrm{\Delta(k)}}$ contains both a 1 and a -1. When analysing whether
partition $j$ requires rebalancing, $\bm{\mathrm{\Delta_j(k)}}$ denotes the
$j$'th column of the matrix, which in turn is the reassignment vector for
partition $j$ at iteration $k$. The column vector illustrated in Eq.
\ref{eq:delta_jk}, would indicate that consumer $|B|$ should stop reading data
from partition $j$, so consumer $1$ can start. It is worth noting that due to
the Bin Packing formulation in Eq. \ref{BPP model} - \ref{opt:c4},
$\bm{\mathrm{\Delta_j(k)}}$ can only have non-zero values in at most 2 rows (2
consumers). 

\begin{equation}
\label{eq:delta_jk}
    \bm{\mathrm{\Delta_j(k)}} = 
        \begin{bmatrix}
            1 \\ 
            0 \\
            \vdots \\ 
            0 \\
            -1
        \end{bmatrix},
\end{equation}

When the column vector only has one non-zero value, if this value is $-1$ then
the consumer group is to stop reading data from this partition at $t_k$ (item
departure). Otherwise, if the value is $1$ then the consumer group has not yet
assigned this partition to the group of consumers, and is to start fetching
messages from this partition at $t_k$ (item arrival). The second scenario
occurs for example at $t_1$ when all partitions have not yet been assigned a
consumer in the group.

\begin{table}[htb!] 
\centering 
\caption{Data required to compute the Rscore.} 
\label{tab:rscore_data}
    \begin{tabular}{ |p{0.05\textwidth}|p{0.4\textwidth}| } 
        \hline 
        \textbf{Notation} & \textbf{Description} \\ 
        \hline 
        $P_k$ & set of partitions to rebalance at time $t_k$ \\ 
        $s_j(t)$ & The write speed of partition $j$ at time $t$ (bytes/s)\\ 
        $C$ & Constant that represents the maximum consumer capacity (bytes/s)\\ 
        \hline 
    \end{tabular} 
\end{table}

Provided the data presented in Table \ref{tab:rscore_data}, Eq. \ref{eq:rscore}
presents the rebalance cost (Rscore) for a single iteration $k$. As presented
in \cite{mann2015approximability}, the migration cost can be related to the
number of migrations performed, or it can be computed based on the total size
of the items being migrated. Rscore considers the latter measure, as it susm
the total production rate into all partitions being migrated.

\begin{equation} 
\label{eq:rscore}
    Rscore(k) = \frac{1}{C}\sum_{j \in P_k} s_j(t_k) \quad \forall \; k \ge 1. 
\end{equation}

In essence, the time it takes to rebalance all partitions, combined with the
Rscore, determines the amount of time a dedicated consumer requires to process
the data that accumulated while migrating the partitions.

To further illustrate what the metric is evaluating, consider that at iteration
$k = 2$ there are 3 partitions being rebalanced with an equal production rate
of $100 bytes/s$, and the consumer's capacity is $100 bytes/s$, then $Rscore_2
= 3$. If it took the consumer group 2 seconds to rebalance, then it would take
a dedicated consumer $3*2 = 6$ full seconds to read the data that has
accumulated while migrating the partitions.

\subsection{Modified Any Fit}
\label{sub:modified_any_fit}

The motivation to modifying the existing Any Fit algorithms, is that the
existing algorithms only focus on reducing the amount of bins used to pack a
set of items, disregarding the cost associated with reassigning items.
Algorithm \ref{alg:modified_any_fit} illustrates the implementation of the
modified any fit algorithms that will be further described in the remainder of
this section. The algorithm describes the procedure to assign partitions to
consumers at time $t_k$ of iteration $k$.

\begin{algorithm}[htb!]
\caption{Modified Any Fit Algorithm}
\label{alg:modified_any_fit}
\begin{algorithmic}[1]
    \renewcommand{\algorithmicrequire}{\textbf{Input:}}
    \renewcommand{\algorithmicensure}{\textbf{Output:}}
\REQUIRE
    current consumer group configuration $C$ \& \\ 
    currently unassigned partitions $U$ \& \\
    $assignStrategy$ (Best or Worst Fit) \& 
    $sortStrategy$ (cumulative or max partition sort)
\ENSURE    
    new consumer group configuration $N$
\STATE {$N \gets $ new ConsumerList with $assignStrategy$} \label{maf:new_consumer_group}
\STATE {$S \gets $ sort $C$ with $sortStrategy$} \label{maf:sort_consumer_group}
\FOR {$c \in S$}
    \STATE $pset \gets $ partitions assigned to $c$
    \STATE $pset \gets $ sort $pset$ in decreasing order \label{maf:sort_partitions}
    \FOR {$i \gets pset.size()-1$ to $0$}
        \STATE $p \gets pset[i]$
        \STATE $result \gets N.assignOpenBin(p)$ \label{maf:assign_open_bin}
        \IF {$result = false$}
            \STATE $break$
        \ENDIF
        \STATE $pset.remove(p)$
    \ENDFOR
    \IF {$pset.size() = 0$}
        \STATE $continue$
    \ENDIF 
    \STATE $N.createConsumer(c)$ \label{maf:create_consumer}
    \FOR {$p \in pset$} \label{maf:start_assign_current_consumer}
        \STATE $result \gets N.assign(c, p)$
        \IF {$result = false$}
            \STATE $break$
        \ENDIF 
        \STATE $pset.remove(p)$
    \ENDFOR \label{maf:end_assign_current_consumer}
    \STATE $U.extend(pset)$ \label{maf:extend_unassigned}
\ENDFOR
\STATE $U \gets$ sort $U$ decreasing order \label{maf:start_final_stage}
\FOR {$p \in U$}
    \STATE $N.assignBin(p)$
\ENDFOR \label{maf:end_final_stage}
\RETURN $N$
\end{algorithmic}
\end{algorithm}

Given the current consumer group's state (the partitions assigned to each
consumer), and a set of unassigned partitions, the modified algorithms differ
from the decreasing versions of the Any fit algorithms described in Section
\ref{sub:approximation_algorithms}, wherein the former class of algorithms
sorts the consumers of a consumer group instead of the set of items. 

At the beginning of the algorithm, an empty consumer group is initialized,
which will modified and returned at the end of the iteration to indicate the
future consumer-partition assignments.

First, the consumers (bins) in the current consumer group are sorted from
biggest to smallest based on two strategies (Line
\ref{maf:sort_consumer_group}): Sort each consumer based on the cumulative
speed of all partitions assigned to it (cumulative sort); Sort each consumer
based on the partition assigned to it that has the biggest measured write speed
(max partition sort). The sorting strategies present some similarities to the
\textit{ServerLoad} algorithm presented in \cite{shi2013empirical}, however
their algorithm process the bins from smallest to biggest. The remainder of our
algorithm also differentiates our procedure.

After sorting the consumer group configuration from iteration $k-1$ using one
of the above strategies, for each consumer in the sorted group, the partitions
assigned to it are sorted based on their write speed (Line
\ref{maf:sort_partitions}). From smallest to biggest, each partition is
inserted into one of the bins that has already been created in the new consumer
group initialized at the beginning. Assigning the items from smallest to
biggest focuses on migrating the partitions with lowest rates of production,
which in turn reduces the migration cost. If the insert is successful, then the
partition is removed from the sorted list of partitions, else (i.e. there is no
existing bin that can hold the partition) then the current consumer assigned to
the partition, is created (Line \ref{maf:create_consumer}). We obtain this
consumer via the index of the row in the column vector
$\bm{\mathrm{x}_j(t_{k-1})}$ which presents the value $1$.

The remaining partitions in the sorted list are now inserted into the newly
created bin, from biggest to smallest (Lines
\ref{maf:start_assign_current_consumer} -
\ref{maf:end_assign_current_consumer}). The current consumer had to be created
since the partitions couldn't all be migrated, therefore, we now try to assign
the biggest partitions first so their production rate does not contribute to
this iteration's rebalance cost. If a partition is inserted successfully, it is
removed from the list of partitions. If a partition does not fit into the
consumer it is currently assigned, then the remaining partitions are added to
the set of unassigned partitions (Line \ref{maf:extend_unassigned}). 

After performing the same procedure over all consumers, there is now a set of
partitions which have not been assigned to any of the consumers in the future
assignment. The final stage (Lines \ref{maf:start_final_stage} -
\ref{maf:end_final_stage}) involves first sorting the unassigned partitions in
decreasing order (based on their measured write speed), and each partition is
assigned a consumer from the new consumer group using the defined fit strategy.

The proposed algorithms that aim to improve the existing Bin Packing heuristics
with respect to the rebalance cost, are obtained through all combinations of
the aforementioned sort and fit strategies, summarised in Table
\ref{maf:table}.

\begin{table}[htb!] 
\centering 
\caption{Modified implementations of the any fit algorithms.} 
\label{maf:table}
\begin{tabular}{ |p{0.4\columnwidth}|p{0.5\columnwidth}| } 
    \hline 
    \textbf{Algorithm} & \textbf{Consumer Sorting Strategy} \\ 
    \hline
    Modified Worst Fit (MWF) & Cumulative write speed \\ 
    \hline
    Modified Best Fit (MBF) & Cumulative write speed \\ 
    \hline
    Modified Worst Fit Partition (MWFP) & Max partition write speed \\ 
    \hline
    Modified Best Fit Partition (MBFP) & Max partition write speed \\
    \hline
\end{tabular} 
\end{table}

\section{Kafka Consumer Group Autoscaler}
\label{sec:autoscaler}

\begin{figure*}[htb!] \centering
    \includegraphics[width=0.9\textwidth]{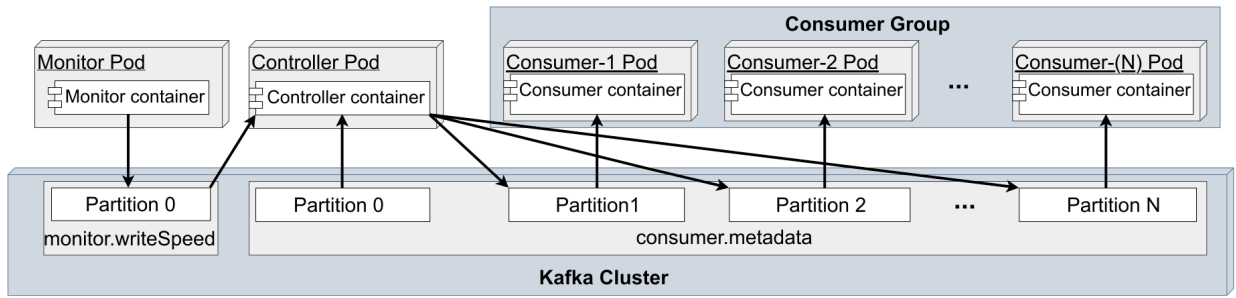}
    \caption{System's architecture based on Kafka consisting of Monitor, Controller and Consumer processes.} 
    \label{fig:system_architecture}
\end{figure*}

In this section, we consider Kafka as the message broker for evaluation purposes, although the same methodology can be applied to other message broker systems. There are three components, presented in Fig. \ref{fig:system_architecture}, that interact with one another to model the BP problem, in order to provide a fully dynamic pipeline capable of autoscaling
based on the current load of data being produced to the partitions of interest.

The monitor process is responsible for measuring the write speed of each
partition the group is interested in consuming data from, which is equivalent to
specifying the size of the items of the bin packing problem. This information is then delivered
to the controller, which is responsible for managing a consumer group (creating
and deleting consumer instances), and mapping each partition to a consumer. The
consumers are then informed of their tasks and read the data from the partitions
that were assigned to them by the controller.

This system was applied in a real world context, with the goal of extracting data published to Kafka topics to be inserted into a Datalake (source code\footnote{\href{https://github.com/landaudiogo/kafka-consumer-group-autoscaler}{https://github.com/landaudiogo/kafka-consumer-group-autoscaler}}).

\subsection{Monitor}
\label{sec:autoscaler:monitor}

To solve the BP problem, initially the controller requires as input the write
speed of each partition the consumer group is interested in consuming data from.
The monitor process is responsible for providing this information.

Kafka provides an Admin client, which can be used to administer the cluster, and
also query information about it. This client/class exposes a method
$describeLogDirs()$ which queries the cluster for the
amount of bytes each TopicPartition has. A TopicPartition is a string-integer
pair, which identifies any partition (integer) within a topic (string). 

Each time the partition size is queried by the admin client, a timestamp is
appended to the measurement, and it is inserted to the back of a queue. Any
query that is older than $30$ seconds, which is guaranteed to be in the front of the
queue, is removed. To obtain the write speed of a single partition, the last
element of the queue and the first (representing the latest and the earliest
measurement of the partition size within the last $30$ seconds) are used to compute the
ratio between the difference in bytes and the difference in time ($bytes/s$).
This is also the average write speed over the last 30 second time window. The smaller the time window, the closer the measurement reflects a producer's network capacity as opposed to measuring the current rate of production into a partition. Also, with a small time window, 2 consecutive events arriving close to the same time, will also lead to misleading and noisier production rate measurements. Having evaluated several different sized windows, 30 seconds provides with an acceptable trade-off between the amount of noise and the time it takes for the measurement to converge to a stable production rate.

After computing the write speed for all the partitions of interest, the
information has to be communicated to the controller. To benefit
from an asynchronous approach, this monitor process communicates with the
controller process via a Kafka topic illustrated in Fig. \ref{fig:system_architecture} as $monitor.writeSpeed$. 

\subsection{Consumer}
\label{sub:consumer}

\begin{figure}[!htb] 
    \centering
    \includegraphics[width=0.35\textwidth]{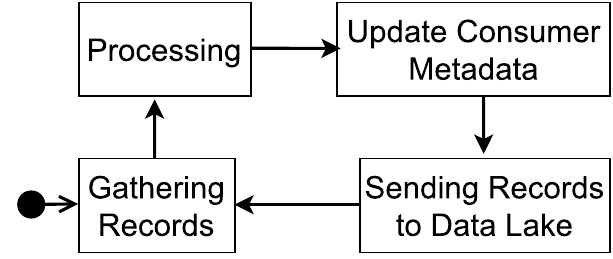}
    \caption{Consumer Insert Cycle.}
    \label{fig:consumer_cycle} 
\end{figure}

The Consumer goes through four important phases within its process, to
approximate its consumption rate to a constant value when being challenged to
work at its peak performance. These phases repeat cyclically until the consumer
is terminated by an external termination signal. Fig. \ref{fig:consumer_cycle}
illustrates the consumer's process.

The consumer is configured with two important parameters,
\textsc{\MakeLowercase{BATCH\_SIZE}} and \textsc{\MakeLowercase{WAIT\_TIME\_SECS}}, which indicate
respectively, the amount of bytes the consumer waits to gather in a single
iteration, and the amount of time it is allowed to wait to gather the
information. 

The first phase has the consumer attempt to fetch \textsc{\MakeLowercase{BATCH\_SIZE}} bytes from
the partitions it is consuming data from. After this condition is satisfied or
\textsc{\MakeLowercase{WAIT\_TIME\_SECS}} is exceeded, the consumer moves on to the second
phase. Here the consumer deserializes and processes each individual record, and batches records
that originated from the same topic together. There are different data lake
tables for each topic, and therefore not all records are inserted into the same
destination table. The number of topics impacts the number of batches prepared, since each table has a dedicated set of batches.

The third phase is where the consumer sends the records into the data lake,
performing an asynchronous request for each topic it fetched data from in the
same iteration.

Having forwarded all the records into their respective tables in the data lake,
the consumer then verifies its metadata queue to verify if there are any change
in state messages to be consumed. This queue is how the Controller process (Sec.
\ref{autoscaler:controller}) informs each consumer of their assignments. If
there are new messages, the consumer reads all messages in the queue, and
updates its state. Only after having successfully updated its state and
persisted its metadata, does the consumer send an acknowledgment back to the
Controller to indicate the successful change in state. This cycle repeats
until the Controller removes the consumer from its group of active consumers. 

\subsection{Controller}
\label{autoscaler:controller}

The controller is the component of the system which is responsible for
orchestrating and managing the consumer group. The write rate of each partition computed
by the monitor process is used as input to the approximation algorithm which
then provides as output a new consumer group configuration. Based on this
configuration, the controller creates the consumers that don't yet exist,
communicates the change in assignment to each consumer in the group, followed by
deleting the consumers that are not required in the new computed group's state.

As shown in Fig. \ref{fig:system_architecture}, there are two topics that
function as communication intermediaries between the three system components.
The topic illustrated by $monitor.writeSpeed$ is where the controller
reads the partitions' write speeds, whereas the topic illustrated by
$consumer.metadata$ is where the controller sends messages to each
consumer to inform the change in their state. Regarding the latter kafka topic,
partition 0 is reserved for communication directed toward the controller,
whereas the remaining partitions within $consumer.metadata$ represent
a one-to-one mapping to each consumer. Therefore, if the controller wants to
communicate with consumer $N$, it sends a record into partition $N$.

This communication architecture was devised to achieve an efficient
communication model, defined as the amount of information that is relevant
relative to the amount of data read. With this pattern, all the data
read is relevant to the reading entity.

\begin{figure}[!htb] 
    \centering
    \includegraphics[width=0.5\textwidth]{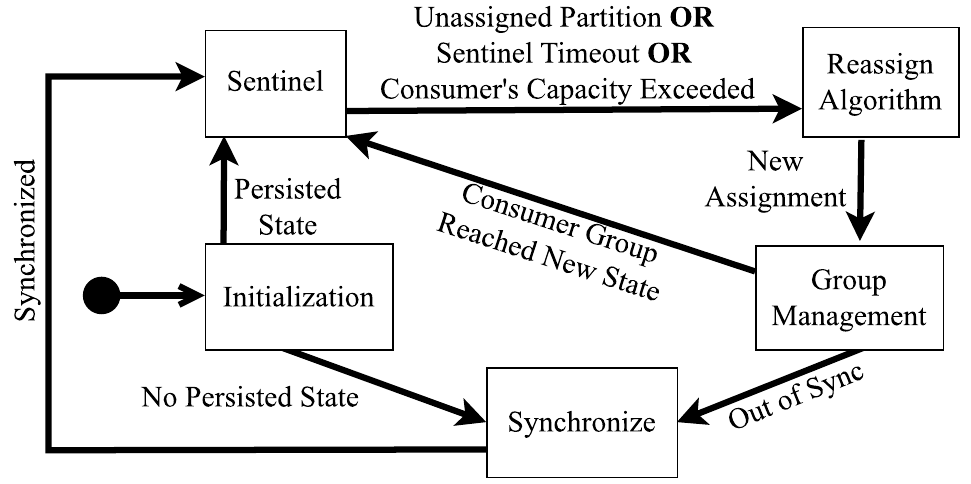}
    \caption{Controller State Machine.} 
\label{fig:state_machine} 
\end{figure}

The controller can be summarized by a state machine, intended to continuously
manage a group of consumers and their assignments, illustrated by Fig.
\ref{fig:state_machine}. 

The Sentinel state has the controller read the information published by the
monitor component to update its current consumer group's state, and determine
whether it has to recompute the group's assignments. If it determines that one
of the exit conditions evaluates to true, the controller transitions into the
Reassign Algorithm state. This is where one of the approximation algorithms
mentioned in Sections \ref{sub:approximation_algorithms} and \ref{sub:modified_any_fit} is employed to solve the Bin Packing
Problem heuristically. 

The output of the approximation algorithm indicates the desired state in which
the controller wants its group of consumers. As such, the first step within the
Group Management state is to compute the difference between the current state of
the group and the desired state. This difference encodes information regarding:
the consumers to be created; the partitions each consumer has to stop consuming
from; the partition each consumer has to start consuming from; and lastly, the
consumers that have no assignment, and therefore can be decommissioned. First, the
controller creates the new consumer instances, which involves the controller
communicating with a kubernetes cluster to create the deployments related to
each consumer. We chose to have different deployments for each consumer, so that
the information related to which $consumer.metadata$ partition the
consumer had to read from, could be transmitted through the deployment's
manifest $metadata.name$ attribute. Secondly, the controller informs
each consumer of their change in state by transmitting the start and stop
consuming messages to each consumer via their respective
$consumer.metadata$ partition. It is important to note that
communicating the change in state is a synchronous process, to assure that while
rebalancing partitions there is no more than one consumer from the group reading
data from a partition. This means that for each partition being rebalanced, the
controller first sends out the stop consuming message, and only after the
consumer informs the controller of having acted upon the message, can the
controller send out the start consuming record to the new consumer. After the
consumers have reached their intended state, the controller shuts down the
inactive consumers. 

The Synchronize state is used to synchronize the consumer group's real state
with the controller's perceived state. This is to guarantee the controller is
capable of recovering from an unexpected termination, which can lead to an
inconsistent perception of the consumer group's state.
\section{Experimentation}
\label{sec:experimentation}

\subsection{Approximation Algorithms Adaptation}
\label{sub:approximation_algorithms_adaptation}

The existing bin packing heuristic algorithms presented in Sec.
\ref{sub:approximation_algorithms} were developed to solve a single iteration
of the bin packing problem. When considering $t_k, k > 1$, there already exists
a partition to consumer assignment given by $\bm{\mathrm{X}}(t_{k-1})$. 

As the algorithm is executing (assigning partitions to consumers), the list of
consumers currently in use is encoded in $\bm{\mathrm{y}}(t_{k})$. Also as
described in Sec. \ref{sub:approximation_algorithms}, the list of open bins,
are consumers that are already in use in iteration $k$ ($y_i(t_k) = 1, i \in
B$), that can be used to assign additional partitions. Some approximation
algorithms limit the number of open bins, e.g., the NFD algorithm only allows
one open bin, which is also the last bin created.

When these algorithms cannot place partition $j$ into one of the existing open
bins, then a new bin has to be created. The new bin created is commonly
described as the lowest index bin which has not yet been used. Function $f:
\mathbb{N} \to B$, defined in Eq. \ref{eq:lowest_idx_consumer}, given the
current iteration $k$, provides the lowest indexed unused consumer.

\begin{equation}
\label{eq:lowest_idx_consumer}
    f(k) = \min \{i \in B \; | \; y_i(t_k) = 0\}
\end{equation}

There is no guarantee that the bin returned by $f(k)$ is the current bin
assigned to partition $j$ resulting from the consumer group computation at
$t_{k-1}$. That is, by creating the lowest index unused bin and assigning it
the partition, leads to the partition being assigned to a "random" consumer,
which most of the times leads to it being rebalanced, as highlighted in
\cite{song2013adaptive}.

For this reason, we modify the decision of which bin to create, when an item
does not fit any of the existing open bins. To do so, we define $g: P \times
\mathbb{N}_{\backslash \{1\}} \to B$ in Eq. \ref{eq:g}. This function receives
as input a tuple  $(j, k), j \in P \land k \in \mathbb{N}_{\backslash \{1\}}$,
where $j$ is a partition and $k$ the current iteration, and outputs the
consumer assigned partition $j$ from iteration $k-1$.
\begin{equation}
\label{eq:g}
    g(j, k) = i, \; \text{such that}  \; i \in B \; \land \; x_{ij}(t_{k-1}) = 1.
\end{equation}

When creating a new bin in the adapted approximation algorithm's procedure, the
consumer to be started is obtained from the function $h: P \times \mathbb{N} \to
B$ defined in Eq. \ref{eq:h}.

\begin{equation} 
\label{eq:h}
    h(j, k) = \begin{cases}
        g(j, k), & \text{if} \; y_{g(j, k)}(t_k) = 0 \; \land \; k > 1 \\ 
        f(k),    & \text{otherwise}
    \end{cases}
\end{equation}

In other words, if $y_{g(j, k)}(t_k) = 0$ and $k > 1$ (consumer $g(j, k)$ is
not currently in use in iteration $k$ and a consumer group assignment already
exists from a previous iteration $k-1$), then the consumer created is the one
that is currently assigned partition $j$ from iteration $k-1$. Otherwise, the
default strategy is used, which is the lowest index consumer that is currently
not in use from the vector $\bm{\mathrm{y}}(t_{k})$. 

It is important to note that this adaptation does not change the number of bins
created by the existing approximation algorithms. It only impacts the decision
of which unused bins to spawn while executing the bin packing algorithm. This
procedure generally increases the frequency a partition is assigned the same
consumer it was assigned in $k-1$, and therefore reduces the resulting
rebalance cost (Rscore). These adapted approximation algorithms are used as a
baseline to compare with the proposed modified algorithms from Sec.
\ref{sec:algorithm}, with respect to the operational and rebalance cost. 

\subsection{Test Data Generation}
\label{sec:random_data_generation}

To compare the performance of approximation algorithms, it is common to
evaluate them with respect to the algorithms' average-case performance
\cite{kamali2016efficient} and \cite{song2013adaptive}. Similar to the
evaluation procedure described in \cite{kamali2016efficient}, we create streams
of production rate measurements wherein from one measurement to the next, the
measured production might vary up to 25\% of a consumer's consumption capacity.
However instead of only 5 consecutive measurements, we have 500, and instead of
also focusing on the departure and arrival of items, our focus lies on the
items changing sizes between 2 consecutive measurements.

Throughout this section, we use the term measurement to represent a map of
values that indicate the speed for each partition of interest, and the term
stream to represent a sequence of measurements in the form of a list. Each
measurement of a stream simulates the items' sizes for an instance of the BP
problem, which is to be fed as input to the approximation algorithms. 

\begin{table}[H] 
\centering 
\caption{Data to generate a stream.} 
\label{table:testing_data} 
\begin{tabular}{ |c|l| } 
    \hline 
    \textbf{Symbol} & \textbf{Description} \\ 
    \hline 
    $P$ & Set of partitions of interest for the consumer group. \\ 
    $s_j(t_k)$ & Speed for a partition $j \in P$ at iteration $k$. \\ 
    $\phi(\delta)$ & Uniform random function that selects a value \\
                   & between $[-\delta, \delta]$. \\
    $C$ & Bin Capacity for the Bin Packing Problem.  \\
    \hline 
\end{tabular} 
\end{table}

To generate a stream of measurements, given $N$ (the number of measurements
desired) and $\delta$ (maximum relative speed variation between two sequential
iterations), at first the initial speed $s_j(t_1), \; \forall j \in P$ has to
be defined. Four different approaches were tested for the partition's initial
value: choosing a random value between $[0, 100]\% \cdot C$; setting the
initial speed of all partitions to $0$; setting the initial speed of all
partitions to $50\% \cdot C$; setting the initial speed of all partitions to
$C$. Given that there was no significant difference on the outcome with these
variations, the results presented were obtained using the streams of data
generated having an initial speed for all partitions that was randomly selected
between $[0, 100]\% \cdot C$.

Therefore, given $s_j(t_1)\ \forall j \in P$, the remaining measurements were
obtained using: 
\begin{equation}
\begin{split}
    s_j(t_k) = &  \max\{
                    0, 
                    s_j(t_{k-1}) + \frac{\phi(\delta)}{100} \cdot C
                \}, \\
             & \forall \ j \in P \ \wedge \ 
                         i \in \{1, 2, ..., N\}
\end{split}
\end{equation}

Using the aforementioned procedure, 6 different streams of data were generated
by setting $N = 500$ and setting $\delta$ to a value belonging to the set $\{0,
5, 10, 15, 20, 25\}$ for each stream of data ($\delta$ does not change within a
stream), which respectively indicate the number of consecutive measurements in
the stream and the variability factor between two consecutive iterations.

\subsection{Comparison with Kafka's Assignment Strategy}

This section concerns itself with evaluating whether rebalancing partitions
between the consumers in a group in a load-aware manner is more beneficial with
respect to latency compared to simply splitting the load by assigning an equal
number of partitions to each consumer. This is a direct comparison between our
proposed approach from Sec. \ref{sub:modified_any_fit} and solutions that
leverage Kafka's assignment strategy such as KEDA. Throughout this section we
consider the following assumptions: An iteration has a duration of $30s$; a new
measurement is read every $30s$; the production rate is stable throughout a
whole iteration $k$; it takes $5s$ to rebalance partitions.

The following example provides a simplified model we use to measure the
experienced latency by the data produced to a queue. Let's consider a consumer
has a consumption rate of $10 bytes/s$, and it has been assigned two
partitions, each with a production rate of $8 bytes/s$. Our consumer reads a
byte every $\frac{1}{10} = 0.1s$ and a new message is ready to be read every
$\frac{1}{8+8} = 0.0625s$. If the consumer is ready to read a new message but
there is no message to be read, when a message does arrive we consider it to be
read with 0 latency. This is the case for the first byte sent to the queue. The
next byte is sent after $0.0625s$ seconds, however the consumer is only ready
to read the next byte after $0.1s$, which implies a latency of $0.0375s$. The
byte following that will be sent after $0.125s$, and the consumer will only
read it after $0.2s$, which corresponds to a latency of $0.075s$, and so on for
the remaining data that is produced at this rate. For this consumer, its
latency model could be expressed by $l = (0.1-0.0625)x$. This linear equation
is valid while the rate of production the consumer has to keep up with is $8+8
bytes/s$.

To express the latency experienced by configurations that leverage Kafka's
assignment strategy, we first have to understand how it performs the
partition-consumer assignment. Given a number of consumers, Kafka assigns an
approximately equal number of partitions to each consumer. If there are $20$
consumers and $32$ partitions, $12$ of them would be assigned $2$ partitions
and $8$ would be assigned $1$ partition ($32$ partitions total). The
partition-consumer assignment is random, and there are no guarantees that the
total rate of production into each consumer does not exceed its consumption
capacity (condition that leads to increased latency). In this case, all latency
measured is a consequence of the difference between the total assignment
production rate and the consumer's consumption capacity $\bar{C}$.

On the other hand, the bin packing algorithm procedures described in Sec.
\ref{sub:modified_any_fit} guarantee that the rate of production never exceeds
the consumer's rate of consumption, leading to no latency in stable conditions.
However, to comply with this condition, as the production rate into each
partition varies, it might be necessary to rebalance partitions from one
consumer to another, a procedure which is not performed by Kafka's assignment
strategy if there is no resizing. Throughout the whole rebalance procedure,
which we compute to be $5s$ in our implementation in Sec. \ref{sec:kafka_cga},
the data produced into the partitions being rebalanced cannot be consumed,
i.e., only after the rebalancing terminates may the consumer start reading the
data from its newly assigned partitions. For this reason, any latency measured
in these algorithms is a consequence of this migration.

The difference between both types of latencies requires we consider 2 distinct
queues for each consumer. One queue that contains data produced into the
partitions rebalanced in the considered iteration, and another for data
produced into partitions carried over from the previous iteration. The first
type we denote as a \textit{rebalanced} queue and the second as a
\textit{fixed} queue.

Data produced into the rebalanced queue cannot be consumed until after $5s$
since the start of a new iteration, which corresponds to the time to rebalance
the partitions. Conversely, a consumer is always consuming data from the fixed
queue. As an example of the rebalanced queue's case, if there are 2 partitions
being migrated to a consumer each with a production rate of $8bytes/s$, then
throughout the whole iteration ($30s$) these partitions would write $480 bytes$
($16 \cdot 30$) into the rebalance queue. However, the consumer only has
$30-5=25s$ to consume all the data written to this rebalanced queue. 

Furthermore, each consumer is reading data from both their fixed and rebalanced
queues. Consequently, the sum of the rate at which data is read from the fixed
queue and from the rebalanced queue, may never exceed the consumer's maximum
consumption rate $\bar{C}$. Nevertheless, in our bin packing formulations (Sec.
\ref{sub:modified_any_fit}), we refer to another capacity $C$, which encodes
the maximum total production rate that can be assigned a single consumer. This
difference is convenient for the bin packing algorithms to allow a consumer to
catch up with the messages produced throughout the rebalance procedure.

To formulate the difference required between $C$ and $\bar{C}$, we have to
consider the worst rebalance case for a single consumer. This occurs when the
sum of the write rate of the partitions being rebalanced is equal to $C$, the
bin's capacity (it may never exceed C due to the bin packing capacity
condition). Hence, the total amount of bytes produced into the rebalanced queue
throughout a single iteration is $30 \cdot C$. Considering the consumer only
has $25s$ to consume the data produced to the rebalanced queue, we now indicate
how we should compute the bin's capacity $C$ based on the consumer's real
capacity $\bar{C}$ to guarantee that the rebalanced queue latency reaches $0s$
within a single iteration. For this to be the case, $30 \cdot C < 25 \cdot
\bar{C}$, which leads to $C < \frac{5}{6} \bar{C}$. This provides the
bin's capacity to use in the bin packing algorithms based on the consumer's
real capacity $\bar{C}$. The drawback to specifying a lower $C$ is that it
leads to an increase in the number of consumers deployed (operational cost), as
the total production rate we can fit into a bin is now smaller than the
consumer's real capacity ($\bar{C}$).

The partitions rebalanced to a consumer $c$ in iteration $k$ is computed by Eq.
\ref{eq:partitions_rebalanced}. 
\begin{equation}
\label{eq:partitions_rebalanced} 
    P_R^k(c) = \{ j \; | \; \Delta_{cj}(k) = 1 \; \land \; j \in P \}. 
\end{equation} 
Condition $\Delta_{cj}(k) = 1$ guarantees a partition is assigned to consumer
$c$ at iteration $k$ but wasn't at $k-1$ (Sec. \ref{sec:problem_formulation}).
Set $P_R^k(c)$ is built upon $P$, which is the set of all partitions the group
is interested in consuming data from.

On the other hand, Eq. \ref{eq:partitions_carry} specifies the partitions which are
carried over from the previous iteration
\begin{equation}
\label{eq:partitions_carry}
    P_F^k(c) = \{ j \; | \; \bm{\mathrm{X}}_{cj}(k) = 1 \; \land \; j \not\in P_R^k \; \forall \; j \in P \},
\end{equation}
wherein $\bm{\mathrm{X}}_{cj}(k) = 1$ makes sure the partition is assigned to
consumer $c$ in iteration $k$, with the additional condition that $j$ does not
belong to the set of rebalanced partitions computed by Eq.
\ref{eq:partitions_rebalanced}.

The consumer's fixed queue contains the data produced into partitions
partitions ($P_F^k(c)$), and the rebalanced queue the data produced into
partitions partitions ($P_R^k(c)$). The rate at which data is produced into the
rebalanced and fixed queues is computed as the sum of each partition's write
rate. Hence, the rate of production into the fixed queue is expressed by Eq.
\ref{eq:production_rate_fixed}, and the rate at which data is produced into the
rebalanced queue expressed by Eq. \ref{eq:production_rate_rebalanced}. 

With regards to the rate at which the consumer reads data from each queue, if
the production rate into the rebalanced queue $W_R^k(c) = 0$ then the speed at
which the consumer reads data from the fixed queue is $R_F^k(c) = \bar{C}$,
otherwise, $R_F^k(c) = min(\bar{C}, W_F^k(c))$. I.e. The consumer reads from
the fixed queue with its full capacity if there are no rebalanced partitions or
the rate of its consumption perfectly matches the rate of production into this
queue (limited to its capacity). If the consumer has newly assigned partitions
(rebalancing), which happens in our bin packing algorithms, all excess
consumption throughput is dedicated to the rebalance queue and therefore the
rate at which the consumer reads from the rebalance queue is $R_R^k(c) =
\bar{C} - R_F^k(c)$. 

Lastly, to express the latency equation for the fixed queue, we define two
additional parameters specified by Equations \ref{eq:total_bytes_rebalanced}
and \ref{eq:total_bytes_fixed}, which respectively compute the total amount of
bytes produced into the rebalanced and fixed queues at an iteration $k$ for a
consumer $c$. We then use Equations \ref{eq:latency_fixed} and
\ref{eq:latency_rebalanced} to measure the latency on the fixed and rebalanced
queues respectively.

\begin{alignat}{3}
    &W_F^k(c) = \sum_{j \in P_F^k(c)}{s_j(t_k)} \label{eq:production_rate_fixed}\\
    &m_F^k(c) = \left ( \frac{1}{R_F^k(c)} - \frac{1}{W_F^k(c)} \right )\\
    &b_F^k(c) = 
        \begin{cases}
            L_F^{k-1}(c, T_F^{k-1}(c)), \quad if \; k > 1\\
            0, \quad otherwise
        \end{cases}\\
    &T_F^k(c) = 30 \cdot W_F^k(c) \label{eq:total_bytes_fixed}\\
    &L_F^k(c, i) = max \left \{ m_F^k(c) \cdot i + b_F^k(c), 0 \right \} \quad && \forall i \in T_F^k(c) \label{eq:latency_fixed}\\
\end{alignat}{3}

To evaluate each algorithm over the same measurements, we first generate a
stream of measurements through the procedure described in Sec.
\ref{sec:random_data_generation} with $32$ partitions, $N=100$ measurements and
$\delta=5$. We assume that each measurement read from the stream indicates a
stable production rate to each partition for the next $30s$. Therefore, for a
stream with $100$ measurements, each experiment has a duration of $100\cdot30 =
3000s$. We then create consumer group configurations based on each algorithm's
assignment strategy and compute the latency as per the models defined in
Equations \ref{eq:latency_fixed} and \ref{eq:latency_rebalanced}.

\begin{alignat}{3}
    &W_R^k(c) = \sum_{j \in P_R^k(c)}{s_j(t_k)} \label{eq:production_rate_rebalanced}\\
    &m_R^k(c) = \left ( \frac{1}{R_R^k(c)} - \frac{1}{W_R^k(c)} \right )\\
    &T_R^k(c) = 30 \cdot W_R^k(c) \label{eq:total_bytes_rebalanced}\\
    &L_R^k(c, i) = max \left \{ m_R^k(c) \cdot i + 5, 0 \right \} \quad        && \forall i \in T_R^k(c) \label{eq:latency_rebalanced}
\end{alignat}

\begin{figure}[htb!] 
    \centering
    \includegraphics[width=\columnwidth]{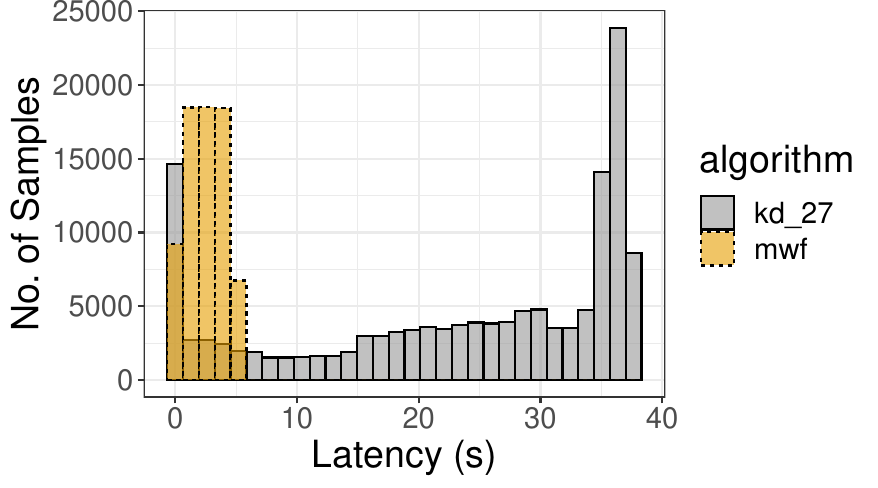}
    \caption{
        Latency histogram for MWF and \texttt{kd\_27}.
    } 
    \label{fig:latency_hist} 
\end{figure}

Figure \ref{fig:latency_hist} provides a histogram comparing Kafka's assignment
for 27 consumers (\texttt{kd\_27}) with our Modified Worst Fit (MWF) algorithm.
Since there are only 32 partitions \texttt{kd\_27}, is 5 consumers away from
the maximum number of consumers allowed for this setup. Despite this fact, it
can be seen that the latency values go up to $40s$. In Kafka's configuration 5
consumers have been assigned 2 partitions. The combined production rate of the
partitions Kafka decides to pair together in a consumer often exceeded the
consumer's consumption capacity $\bar{C}$. As for MWF, only after the rebalance
procedure has terminated may the consumer start reading data again from the
rebalanced queue. Therefore, the first bytes written to this queue will
experience latencies close to $5s$, but because in the worst case there is
always $\bar{C} - C$ of consumption throughput the consumer can dedicate to the
rebalanced queue, it can catch up with the messages produced to the rebalanced
queue before the end of an iteration $k$.

\begin{figure}[htb!] 
    \centering
    \includegraphics[width=\columnwidth]{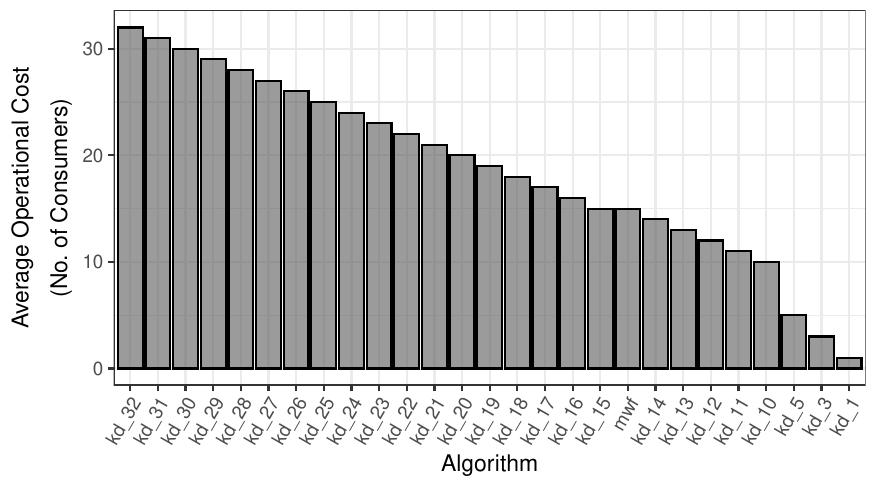}
    \caption{
        Average operational cost for all evaluated algorithms.
    } 
    \label{fig:latency_operational_cost} 
\end{figure}

How do these algorithms compare with respect to their operational cost? Figure
\ref{fig:latency_operational_cost} shows the operational cost of all evaluated
algorithms computed as the average number of consumers required throughout the
whole experiment. Since the Kafka algorithms (\texttt{kd\_1}, \texttt{kd\_3},
\texttt{kd\_5}, etc...) do not resize throughout the experiment, their average
operational cost is equal to the integer value in their name, e.g.,
\texttt{kd\_25} used 25 consumers. In the case of MWF, it used an average of 15
consumers throughout the whole experiment.

\begin{figure}[htb!] 
    \centering
    \includegraphics[width=\columnwidth]{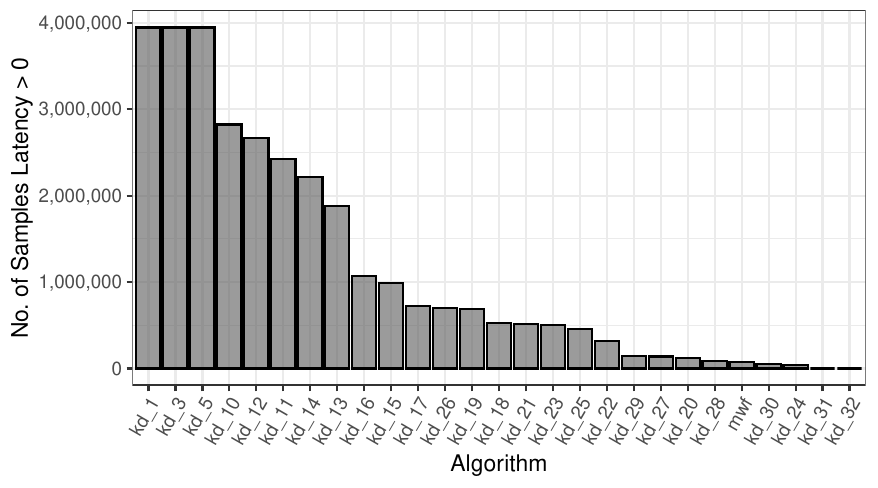}
    \caption{
        Total samples with latency greater than $0s$ for each algorithm.
    } 
    \label{fig:latency_samples} 
\end{figure}

\begin{figure}[htb!] 
    \centering
    \includegraphics[width=\columnwidth]{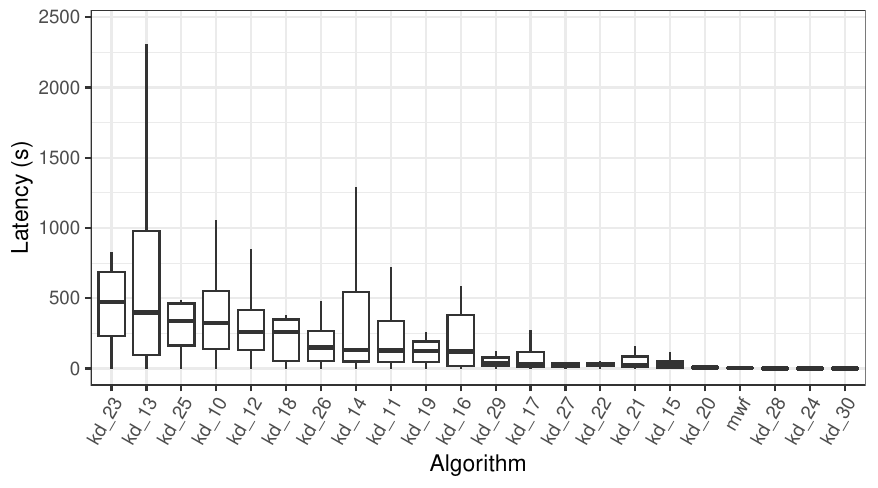}
    \caption{
        Latency samples greater than $0s$ boxplot per algorithm.
    } 
    \label{fig:latency_boxplot} 
\end{figure}

For each algorithm, Figure \ref{fig:latency_samples} plots the number of
samples measured throughout an experiment where the measured latency is above
$0s$. This plot is to be seen in combination with Figure
\ref{fig:latency_boxplot} which shows the boxplot for the latency measurements.
Configurations \texttt{kd\_31} and \texttt{kd\_32} are not present in the
boxplot as they present no samples with latencies above $0s$. Figure
\ref{fig:latency_samples} shows MWF is only outperformed in terms of the total
amount of latency samples greater than $0s$ by Kafka configurations with $24$,
$30$, $31$ and $32$ consumers. MWF presents a $90^{th}$ percentile latency of
$4.52s$, whereas the aforementioned Kafka configurations present $90^th$
percentile latency values of $2.70s$, $1.11s$, $0s$ and $0s$ respectively. On
the other hand, these same configurations present respectively an increase in
operational cost of $60\%$, $100\%$, $106.67\%$ and $113.33\%$ to MWF's average
use of 15 consumers. Lastly, Kafka's configuration with the same operational
cost as MWF, \texttt{kd\_15}, has as $90^{th}$ percentile latency $217s$, which
is $48$ times the latency shown by MWF.

Conterintuitively, with Kafka's assignment strategy, increasing the number of
consumers in a group does not reliably improve the group's performance with
respect to latency, as shown in Fig. \ref{fig:latency_boxplot}. This is a
consequence of Kafka attempting to assign an equal number of partitions to each
consumer without considering the total production rate it is assigning to each
consumer, which more often than not, exceeds the consumer's capacity.
In other words, this highlights that simply increasing the number of consumers
and leveraging Kafka's assign strategy, as is the case for KEDA's autoscaling
solution, is an unreliable method to improve the group's latency performance.

\subsection{Performance Metrics}

The metrics used to compare the performance between the algorithms are the
Cardinal Bin Score ($CBS_\delta(a)$, Eq. \ref{eq:cbs}) and the Average Rscore
($E(Rscore_a^\delta)$, Eq. \ref{eq:avg_rscore}) over all iterations of each
stream.

\begin{table}[H] 
\centering 
\caption{
    Data to compute the cardinal bin score and the expected Rscore for a stream of measurements.
} 
\label{table:bin_score} 
    \begin{tabular}{ |c|p{0.75\columnwidth}| } 
    \hline 
    \textbf{Symbol} & \textbf{Description} \\ 
    \hline 
    $A$ & Set of the existing and proposed approximation algorithms. \\ 
    \hline 
    $z_a^\delta(i)$ & 
        number of bins used in iteration $i \in \{1, ..., N\}$ of a stream
        defined by $\delta$, by algorithm $a \in A$. \\ 
    \hline 
    $Rscore_a^\delta(i)$ & 
        Rscore for an iteration $i \in \{1, ..., N\}$ of a stream
        defined by $\delta$, by algorithm $a \in A$. \\
    \hline 
\end{tabular} 
\end{table}

The cardinal bin score is calculated using the following expression:
\begin{equation}
\label{eq:cbs}
\begin{split}
    CBS_\delta(a) = 
       &\frac{1}{N}
            \sum_{i=1}^{N} 
                \frac{  
                    z_a^\delta(i) - min_{b \in A} \{z_b^\delta(i)\} 
                }{
                    min_{b \in A} \{z_b^\delta(i)\} 
                }, \\
        &\forall \ a \in A \ \wedge \ \delta \in \{0, 5, 10, 15, 20, 25\}. 
\end{split}
\end{equation}

In Eq. \ref{eq:cbs}, $min_{b \in A} \{z_b^\delta(i)\}$ denotes the minimum
amount of bins used by any algorithm in $A$, for iteration $i$ and the
specified stream identified by $\delta$. The size of the stream of measurements
$N$, is also the number of consumer group configurations computed (bin packing
algorithm executions) in this experimentation.

The CBS expresses how many more bins, on average, an algorithm $a \in A$ has
compared to the algorithm that made use of the least bins, which encodes the
operational cost. The closer this value is to $0$, the more frequently the
algorithm provided the configuration with the least amount of bins, hence
performing better than the other algorithms in $A$ with regard to the
operational cost.

The expected value of the Rscore is used to compare the rebalance cost for a
single stream, and is computed as follows:
\begin{equation}
\label{eq:avg_rscore}
\begin{split}
    E(Rscore_a^\delta) = 
        &\frac{1}{N} 
            \sum_{i=1}^{N} Rscore_a^\delta(i), \quad \\
        &\forall \ a \in A \ \wedge \  \delta \in \{0, 5, 10, 15, 20, 25\}.
\end{split}
\end{equation}
Given Eq. \ref{eq:avg_rscore}, the aim is to also minimize an algorithm's
Average Rscore.

\subsection{Modified Any Fit Evaluation}
\label{sub:exp_modified_any_fit}

Each algorithm was evaluated over a stream of measurements, persisting the
Rscore and the number of bins used for each iteration. The algorithms are
compared with respect to the same stream. The code that compiled the following
results is made available in a public
repository\footnote{\href{https://github.com/landaudiogo/thesis_data}{https://github.com/landaudiogo/thesis\_data}}.

\begin{figure}[htb!] 
\centering
\includegraphics[width=\columnwidth]{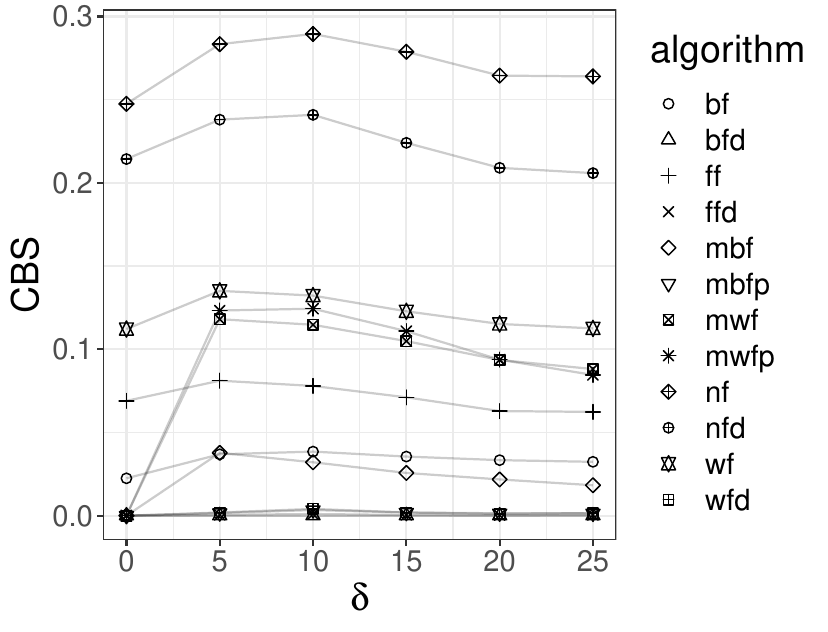} 
\caption{
    Cardinal Bin Score (CBS) for all implemented algorithms.
} 
\label{fig:relative_nconsumers} 
\end{figure}

When evaluating the Cardinal Bin Score, the worst performing algorithm, as
shown in Fig. \ref{fig:relative_nconsumers}, is the next fit followed by its
decreasing version. The remaining any fit decreasing algorithms, are the ones
that perform the best, with the best fit decreasing consistently presenting the
best results. 

\begin{figure}[htb!] 
\centering
\includegraphics[width=0.9\columnwidth]{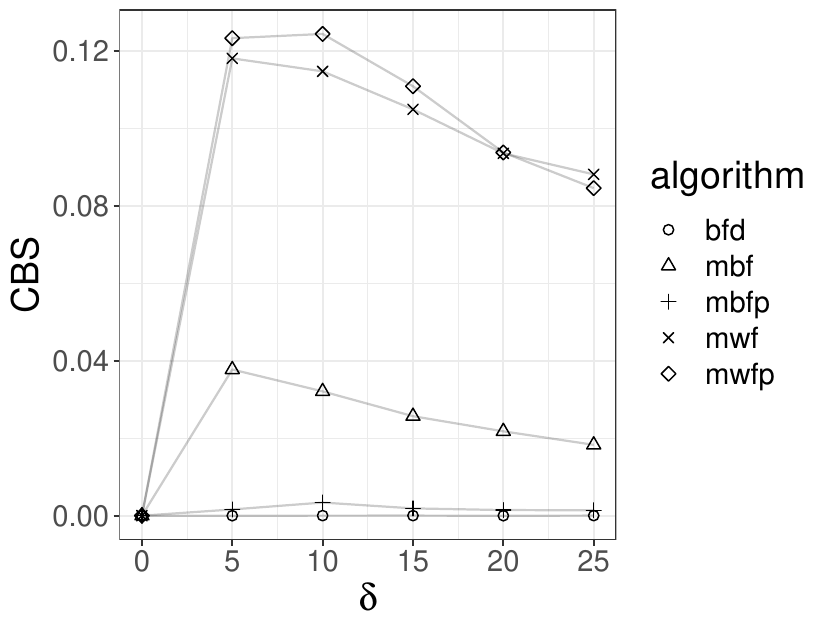} 
\caption{
    Cardinal Bin Score (CBS) filtered to present the modified and the
    BFD algorithms.
} 
\label{fig:relative_nconsumers_modified}
\end{figure}

As for the modified any fit algorithms (Fig.
\ref{fig:relative_nconsumers_modified}), due to its sorting strategy, MBFP
shows the best results. It is also worth noting that for smaller variabilities,
the modified algorithms behave similarly to the online versions of their any
fit strategy with respect to the CBS, since the partitions aren't necessarily
assigned from biggest to smallest. On the other hand, the higher the delta, the
bigger the variability, which also leads to more rebalancing, having the
modified algorithms behave more like the decreasing versions of their fit
strategy (Line \ref{maf:extend_unassigned} and Lines
\ref{maf:start_final_stage} - \ref{maf:end_final_stage} of Algorithm
\ref{alg:modified_any_fit}).

\begin{figure}[htb!] 
    \centering
    \includegraphics[width=\columnwidth]{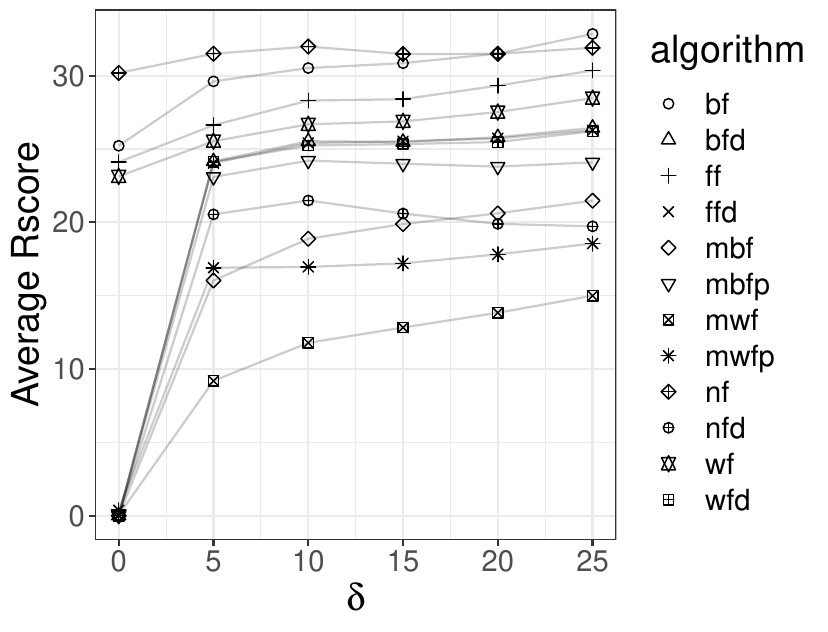}
    \caption{
        Impact on Rscore for different Deltas (random initial partition speed).
    } 
    \label{fig:rscore} 
\end{figure}

With regards to the Rscore, a common trend can be verified for all algorithms,
where an algorithm's Average Rscore increases with the value of $\delta$, as
can be seen in Fig. \ref{fig:rscore}. This should be expected as the bigger the
$\delta$, the more variability there is between the items' sizes in two
consecutive measurements of a stream. 

With respect to the average Rscore, the algorithms that perform the best are
the Modified Any Fit algorithms and the NFD. The reason why the NFD appears
within the five best algorithms, is due to adaptation described in Sec.
\ref{sub:approximation_algorithms_adaptation} and the increased amount of bins
used by this algorithm. The partitions are assigned from biggest to smallest,
and to due the reduced number of allowed open bins in the Next Fit algorithm,
the chances of a partition being assigned to an existing bin are reduced. As
such, the bigger partitions with less likelihood fit within their predecessor's
consumer, and therefore a new bin has to be created. As a consequence of the
adaptation, the new bin created more frequently is the current consumer
assigned to that partition which then leads to an improved rebalance score.

For a similar reason, the modified algorithms that perform the best with
regards to the Rscore, are also the ones that perform worst (compared to the
remaining modified algorithms) when evaluating the CBS. As such, on account of
the added rebalance concern within the modified algorithms, these present an
improvement when it comes to the rebalance cost compared to the existing
approximation algorithms.

\begin{figure}[!htb] 
    \centering
    \subfloat[
        Pareto front for $\delta=5$.
        \label{fig:pareto_front_a}
    ]{
		\includegraphics[width=0.47\columnwidth]{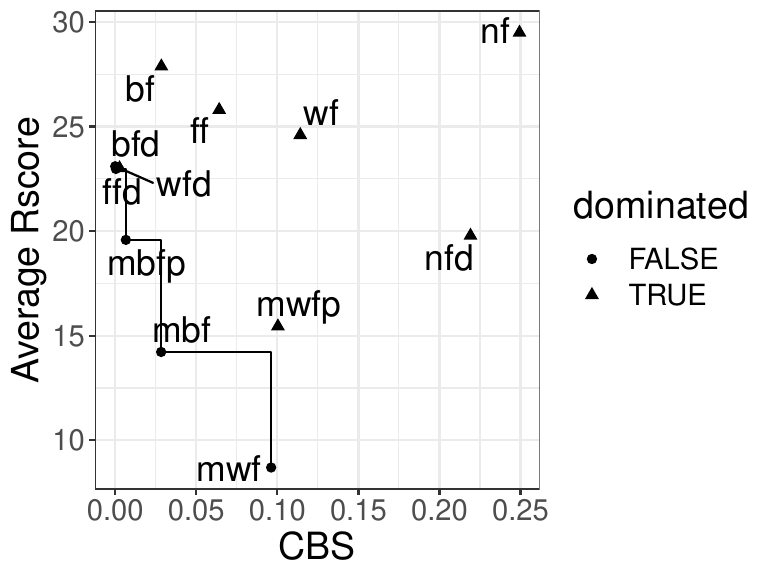}
    }
    \hfill
    \subfloat[
        Pareto front for $\delta=25$.
        \label{fig:pareto_front_b}
    ]{
        \includegraphics[width=0.47\columnwidth]{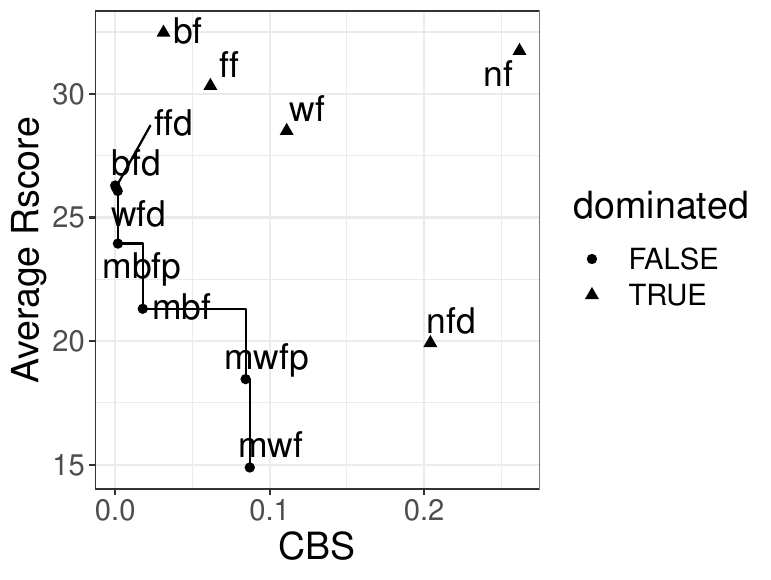}
    }
 		\caption{
        Pareto front for different deltas comparing the Cardinal Bin Score and
        the Average Rscore.
    }
\label{fig:pareto_front} 
\end{figure}

We find the pareto front, so as to determine the set of solutions that are most
efficient, provided there are trade-offs within a multi-objective optimization
problem. Excluding MWFP, the modified algorithms are consistently a part of the
pareto front, as shown in Fig. \ref{fig:pareto_front}, which implies these are
a competitive alternative when solving this variation of the BP problem. For
the tested production rate variation ($\delta$ between 5-25\% of the consumers
consumption capacity), the best performing Modified Any Fit algorithm with
respect to the Rscore is the MWF. For a distributed set of queues that present
a variability up to $25\%$ of a consumer's consumption capacity ($\delta =
25$), MWF reduces the rebalance cost by $23\%$ with an average increase in
operational cost of $8.8\%$. In lower variability setups where the variation in
production rate goes up by $5\%$ of a consumer's capacity, the reduction in
rebalance cost is $55\%$ while only increasing the operational cost by
$11.8\%$.

\subsection{Kafka Consumer Group Autoscaler Evaluation}
\label{sec:kafka_cga}

The consumer group autoscaler, was developed as a proof of concept as to how
this variation of the Bin Packing problem could be applied in a message broker
environment. As such, we now present the results when running this system in a
production environment message broker, with the compute resources utilised
summarized in Table \ref{tab:experimentation_resources}.

\begin{table}[!htb]
\caption{Infrastructure used for experimentation}
\label{tab:experimentation_resources}
\centering
\begin{tabular}{|p{0.18\columnwidth}|p{0.14\columnwidth}|p{0.5\columnwidth}|} 
    \hline
    \textbf{Component} 
        & \textbf{Cloud Platform} 
            & \textbf{Resource Classification}\\
    \hline
    
    Kafka Cluster (3 Nodes) 
        & AWS 
            & r5.xlarge \\
    \hline
    Kubernetes 
        & GCP
            & GKE autopilot \\
    \hline
    Monitor 
        & GCP 
            & Kubernetes Deployment (GKE) \\
    \hline
    Controller 
        & GCP  
            & Kubernetes Deployment (GKE) \\
    \hline
    Consumer Group
        & GCP 
            & As many Deployments as consumers in the group (GKE) \\
    \hline
\end{tabular}
\end{table}

\begin{figure}[htb!] \centering
\includegraphics[width=0.8\columnwidth]{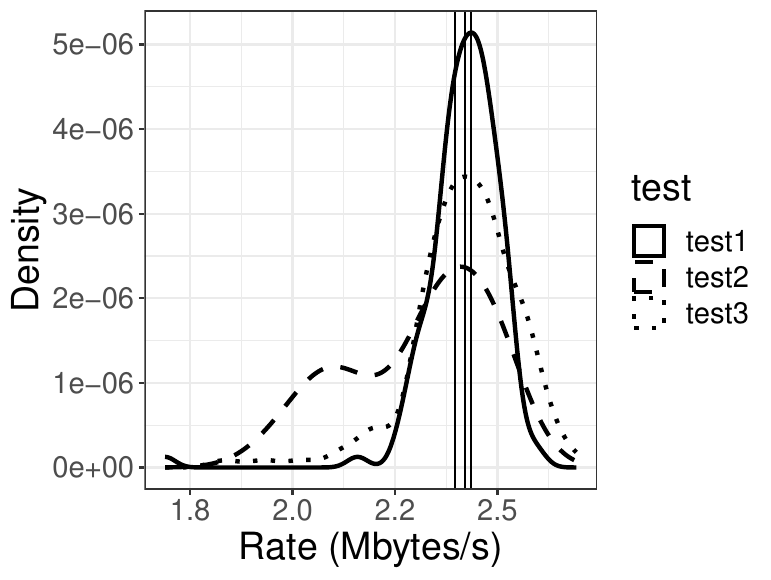}
\caption{
    Density Plot for the consumer's measured throughput in the three testing
    conditions.
} 
\label{fig:consumer_capacity} 
\end{figure}

One of the first assumptions, was that this problem is a variation of the
single bin size bin packing problem, and therefore, the consumer should present
a constant maximum capacity when challenged to work at peak performance. To
validate the assumption, the consumer was tested in three very disparate
conditions, specified in Table \ref{tab:consumer_testing_conditions}, all
requiring the consumer to be reading the data at its maximum consumption rate.
Between each test, we varied: the average amount of bytes present in each
partition; the number of destination tables in the data lake; and the number of
partitions assigned to the consumer instance. Fig. \ref{fig:consumer_capacity}
presents the results of having the consumer run in these conditions, and
clearly shows a common mode for the maximum consumption rate, around $2.3
Mbytes/s$. We therefore consider this assumption to be valid under normal
conditions.

\begin{table*}[htb!] 
\centering 
\caption{
    Testing conditions to obtain consumer maximum throughput measure.
} 
\label{tab:consumer_testing_conditions}
    \begin{tabular}{ |c|r|r|r|r| } 
        \hline
        \textbf{Test ID} & \textbf{Total Bytes} & \textbf{Average Bytes} &
            \textbf{Number of Partitions} & \textbf{Number of Tables} \\ 
        \hline 
        Test 1 
            & $648\ Mbytes$ 
                & $20\ Mbytes$ 
                    & $32$ 
                        & $1$ \\ 
        Test 2 
            & $100\ Mbytes$ 
                & $0.86\ Mbytes$ 
                    & $116$ 
                        & $5$ \\ 
        Test 3 
            & $678\ Mbytes$ 
                & $4\ Mbytes$ 
                    & $144$ 
                        & $5$ \\ 
        \hline
    \end{tabular} 
\end{table*}

With the goal of testing the system as a whole, we statistically summarize the
time it takes the autoscaler to go through each event required to dynamically
scale the group of consumers.  

\begin{figure}[htb!]
    \centering
    \includegraphics[width=0.85\columnwidth]{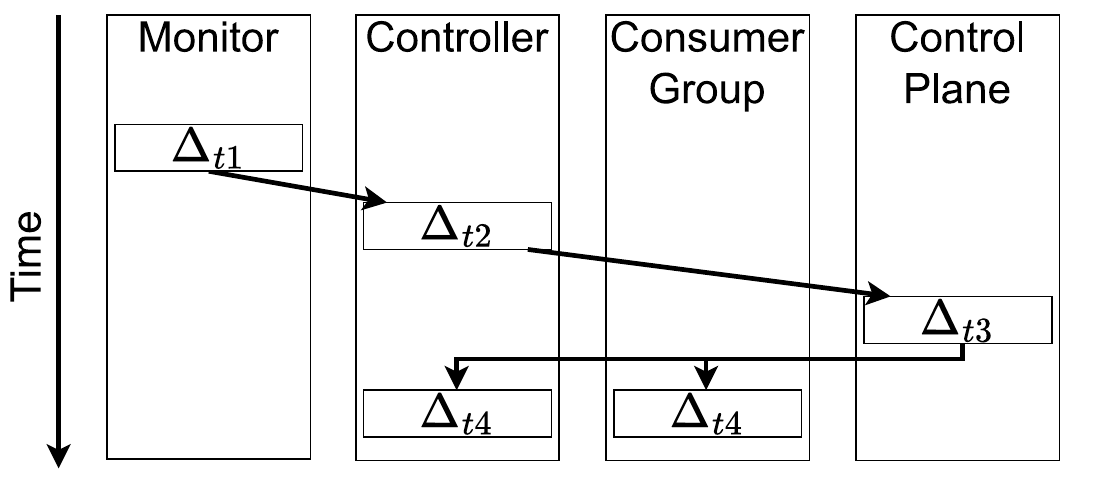}
    \caption{System sequence of response events.}
\label{fig:step_event_sequence}
\end{figure}

Initially, a measurement has to be provided by the monitor process to then be
consulted by the controller ($\Delta_{t1}$). The controller then updates the
group's current state based on the new measurement, and proceeds to computing a
new group configuration using one of the heuristic algorithms presented in
Sections \ref{sub:approximation_algorithms} and \ref{sub:modified_any_fit}.
This is followed by the controller calculating the difference between the new
and current configuration, to determine the consumers it has to create, the
ones it has to delete, and the messages that have to be communicated to each
consumer to reach the intended state (Section \ref{autoscaler:controller}).
This step terminates with the controller communicating with the Kubernetes
control plane, to create the new consumer resources ($\Delta_{t2}$). 

The controller then waits for the newly created deployments to be ready
($\Delta_{t3}$), followed by communicating with the consumer group to inform
the consumers of their change in state, which only terminates as soon as all
messages have been sent out to the respective consumers, and when every message
has been acknowledged back to the controller ($\Delta_{t4}$). This process is
illustrated by Figure \ref{fig:step_event_sequence}.

The system's response time, starts with the monitor process measuring the
current production rate of each partition. The measurement performed by the
monitor component from Sec. \ref{sec:autoscaler:monitor}, is a moving average
with a time frame of 30 seconds. As such, this component would take 30 seconds
to converge to a stable rate of production. This event is represented in Fig.
\ref{fig:step_event_sequence} as $\Delta_{t1}$.

To fully evaluate the system's response to high load variations, and to
increase reproducibility, the monitor process was replaced by the stream of
measurements described in Sec. \ref{sec:random_data_generation}. This provided
more control over the production rate, to analyse the system's response in
scenarios of high and low variability ($\delta = 25$ and $\delta = 5$
respectively) of the production rate.

As for the second event $\Delta_{t2}$, this is the time it takes the controller
to compute the consumer group's new assignment, computing the difference
between the new and current states, and to send an asynchronous request to the
Google Kubernetes Engine (GKE) cluster for every new consumer instance to be
created. To obtain the distribution for this metric, the system was tested with
the streams described in Sec. \ref{sec:random_data_generation}.

\begin{figure}[htb!]
\centering
\includegraphics[width=0.7\columnwidth]{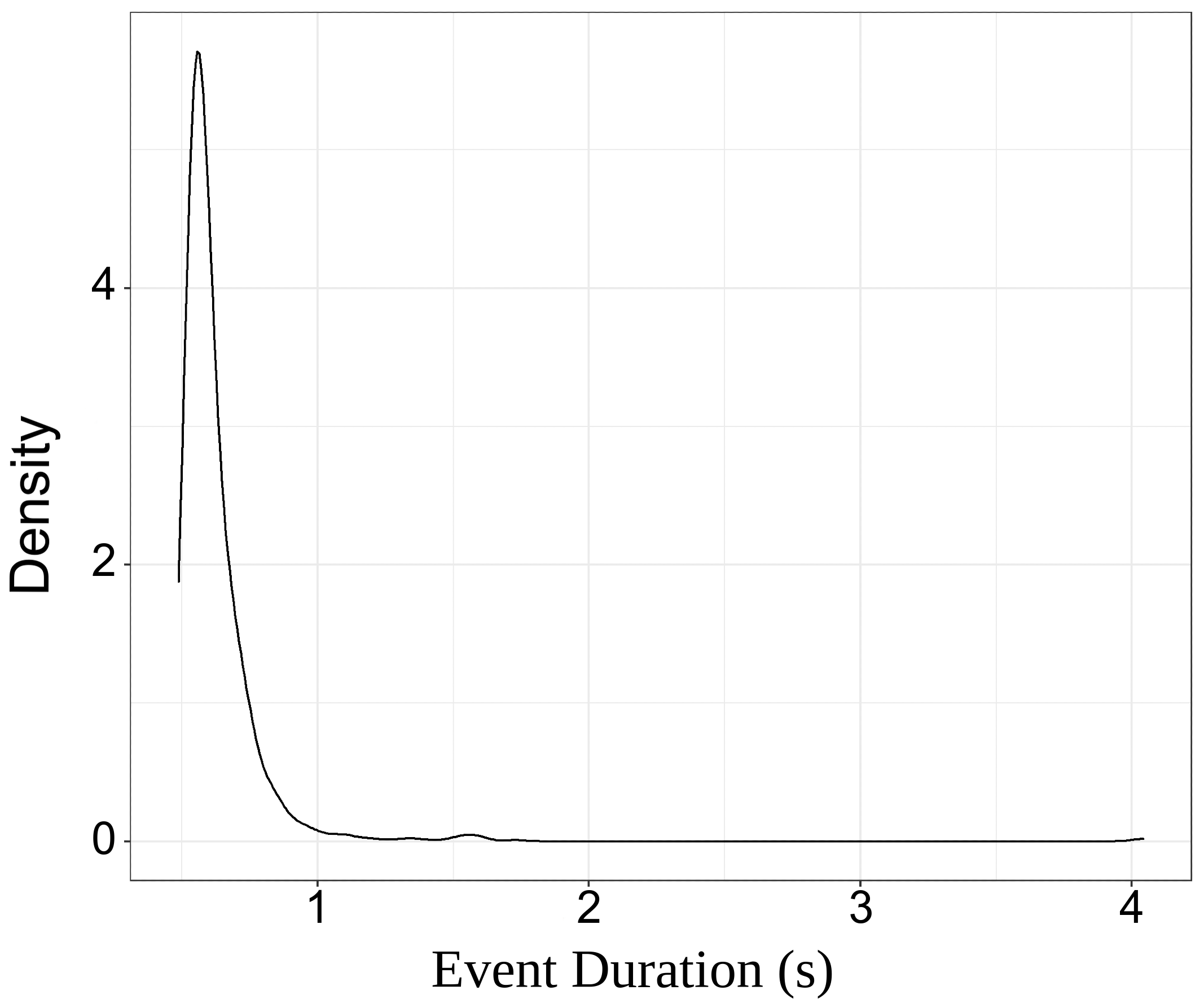}
\caption{
    Distribution of $\Delta_{t2}$ for 1345 observations.
}
\label{fig:controller_result_trigger}
\end{figure}

The time the controller takes with this procedure depends on the number of
partitions to distribute between the consumer group, the algorithm the
controller is executing to figure a new consumer group assignment, and the
number of new consumer instances it has to create in the GKE cluster.

For the tested input data, there were at most 32 partitions to rebalance, at
most 20 consumers to be created in a single iteration and the algorithm
executed was the MWF. The event consistently takes less than 1 second to be
executed as shown in Figure \ref{fig:controller_result_trigger}.

After making the asynchronous request to the GKE cluster, the autoscaler enters
$\Delta_{t3}$, where it has to wait for the control plane to schedule the
consumer pods to cluster nodes, and run the containers.

\begin{figure}[htb!]
\centering
\includegraphics[width=0.7\columnwidth]{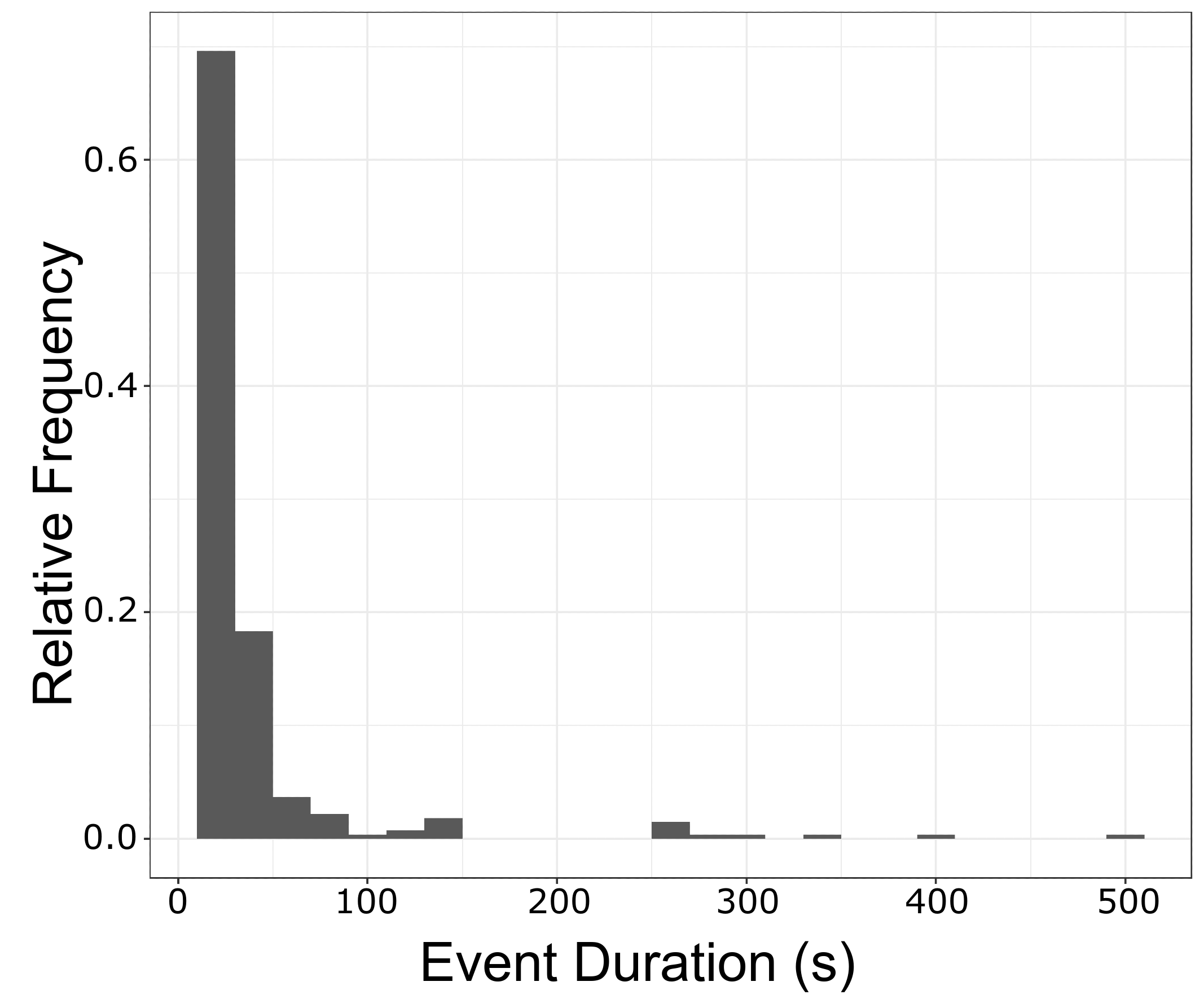}
\caption{
    Histogram of $\Delta_{t3}$ for 273 observations.
}
\label{fig:controller_result_kubernetes}
\end{figure}

From Fig. \ref{fig:controller_result_kubernetes}, there are two main clusters
of data points which can be summarized into two different scenarios in which
the GKE cluster can find itself in.

The first, and most frequent (88\%), has the GKE cluster taking between $[10,
50]$ seconds to have a new consumer instance ready. This is usually the case
when the controller requests for the creation of new consumer instances and the
Kubernetes cluster has enough resources available. As such, the event's
duration is related to the time it takes the scheduled node(s) to download the
image from the container registry, and to start the containers.

The second group of data points, any time span greater than $50s$ (represents
12\% of the data points), occurs when the Kubernetes cluster does not have any
available resources. Here the actions the cluster undergoes are adding a new
node to the cluster, and only then scheduling the consumer instance into the
new available node. Due to the autoscaling feature of the GKE cluster, this is
done automatically but it is also more inconsistent, having data points taking
up to $500s$, although very sporadically.

In spite of the fact there isn't much control over the time it takes the
Kubernetes cluster to schedule and start the pods, one variable that can be
controlled is the size of the image which has to be downloaded by the nodes
that were assigned the newly created consumer instances.

\begin{figure}[htb!]
\centering
\includegraphics[width=0.7\columnwidth]{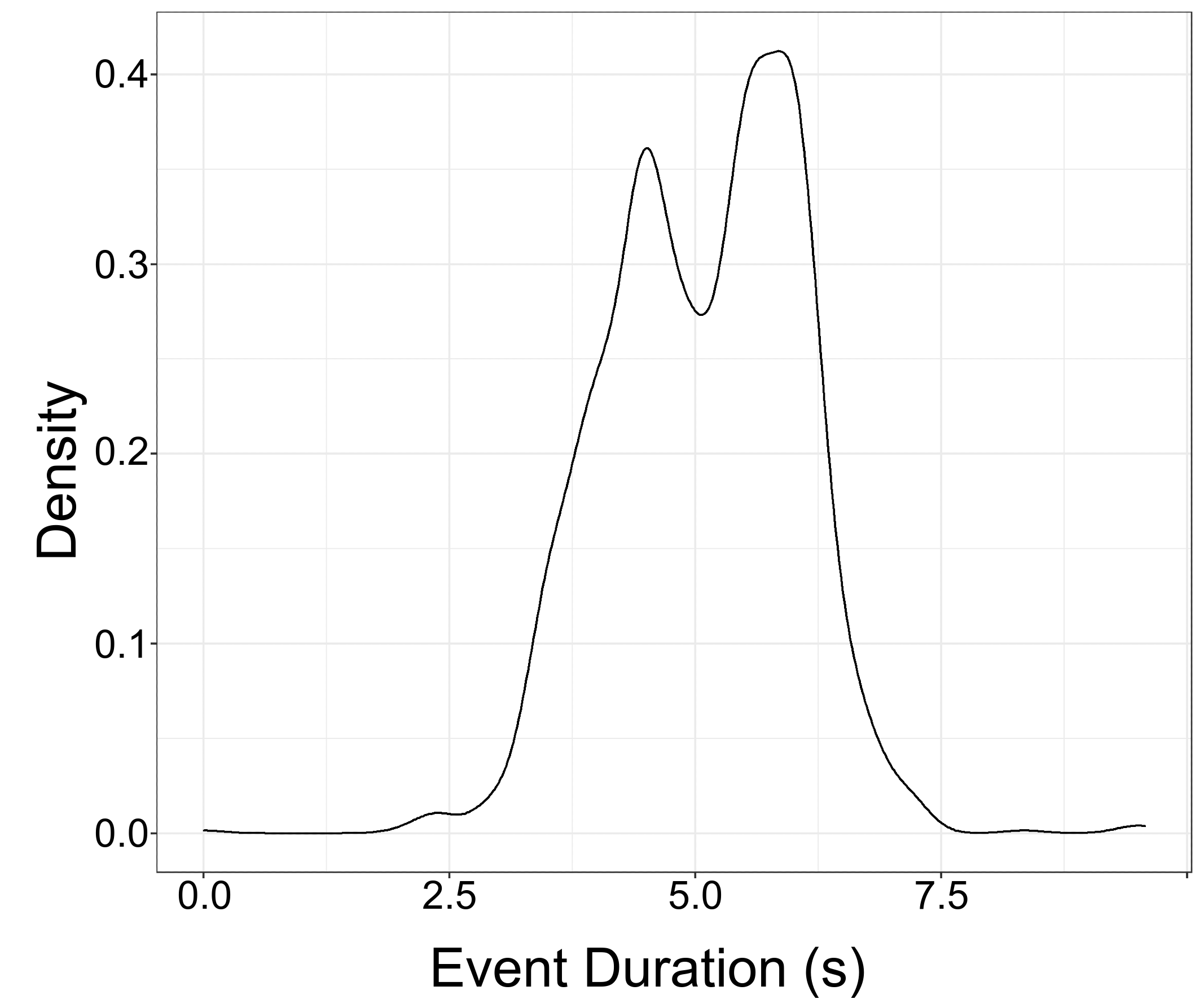}
\caption{
    Distribution of $\Delta_{t4}$ for 1331 observations.
}
\label{fig:controller_result_change_state}
\end{figure}

Lastly, after having the consumer instances running and ready to receive
assignments, the controller has to communicate each consumer's assignment
within the newly computed group configuration, which is represented by
$\Delta_{t4}$ in Fig. \ref{fig:controller_result_change_state}. 

Due to the synchronous nature of rebalancing a partition from one consumer to
another, the controller has to first send out a stop consuming message for each
partition that has to be rebalanced, wait for the consumer to respond, send out
a start consuming message, and wait for a response. Without taking any
processing and network delays into account, at worst, the controller might have
to wait for two consumer cycles to be able to rebalance a partition. 

A consumer's cycle goes through the different phases mentioned in Section
\ref{sub:consumer}, and only after performing it's tasks does it verify the
metadata queue to check if it has received any change in state message. For
this reason, between two metadata reads from the $consumer.metadata$ topic, it
takes the consumer one whole cycle. 

Having defined \textsc{\MakeLowercase{BATCH\_SIZE}} to be $5Mbytes$ and
\textsc{\MakeLowercase{WAIT\_TIME\_SECS}} 1 second, and provided the results
from Figure \ref{fig:consumer_capacity}, which indicate the consumer has a
maximum consumption rate of approximately $2Mbytes/s$, each consumer cycle can
take approximately $2.5s$.  Since the controller has to wait for two consumer
cycles, this would imply that changing the group's state could take around 5
seconds, as can be seen in Fig. \ref{fig:controller_result_change_state}.

It is also worth noting that the more consumers there are in the group, the
higher the probability that the controller has to wait 1 whole cycle after
sending out the stop command, and another whole cycle after a start command, as
the communication would have to be performed with more consumers.

\section{Conclusions}
\label{sec:conclusions}

The work presented in this paper aims to achieve a deterministic approach to
determine the number of consumers required working in parallel so as to
guarantee that the rate of production into a set of partitions is not higher
than the rate of consumption, while minimizing the operational cost.

The goal is to deterministically solve the autoscaling problem related to a
group of consumers, which we model as a Bin Packing problem where the items'
sizes and bin assignments can change over time. In light of these variations,
items' assignments can be rebalanced (bin assignment can change), and therefore
we propose the Rscore to evaluate the rebalance cost, Sec. \ref{sub:rscore}.

Given the conflicting objectives of minimizing both the number of consumers and
the rebalance cost, in Section \ref{sub:modified_any_fit} we propose four new
heuristic algorithms based on the Rscore. We present the Pareto-front for the
evaluated algorithms in Section \ref{sub:exp_modified_any_fit} and show that
three of the proposed algorithms are proven to be competitive solutions when
solving the multi-objective optimization problem. In a message broker
environment with high variability, MWF reduces the rebalance cost (Rscore) by
23\% while increasing the operational cost by 8.8\%. For lower variability
setups, the Rscore is reduced even further to 55\%, while only increasing the
operational cost to 11.8\%.

Furthermore, in Section \ref{sec:autoscaler}, the BP heuristics are applied to
Kafka to automatically scale a group of consumers and manage the
partition-consumer assignments. This system is integrated into a real
production environment, and as expected, is capable of providing a solution
that scales a group of consumers based on the system's load.

Lastly, in this paper, we challenge some of Kafka's core functionalities and
provide tested alternatives that can be used as a foundation so as to improve
the way in which a consumer’s load is modeled and the manner in which the load
(partitions) is distributed between the elements of a consumer group.

\bibliographystyle{IEEEtran}
\bibliography{refs}
\end{document}